# Multi-site benchmark classification of major depressive disorder using machine learning on cortical and subcortical measures

Vladimir Belov[1], Tracy Erwin-Grabner[1], Ali Saffet Gonul[2], Alyssa R. Amod[3], Amar Ojha[4], Andre Aleman[5] ,Annemiek Dols[6] , Anouk Schrantee[7], Aslihan Uyar-Demir[2], Ben J Harrison[8], Benson Mwangi[9,10] , Bianca Besteher[11], Bonnie Klimes-Dougan[12], Brenda W. J. H. Penninx[6], Bryon A. Mueller[13], Carlos Zarate[14], Christopher G. Davey[8], Christopher R. K. Ching[15], Colm G. Connolly[16], Cynthia H. Y. Fu[17,18], Dan J. Stein[19], Danai Dima[20,21], David E. J. Linden[22,23,24,25], David M. A. Mehler[22,23,26], Edith Pomarol-Clotet[27], Elena Pozzi[28,29], Elisa Melloni[30], Francesco Benedetti[30], Frank P. MacMaster[31], Hans J. Grabe[32], Henry Völzke[33], Ian H. Gotlib[34], Jair C. Soares[9,10], Jennifer W. Evans[35], Kang Sim[36,37,38], Katharina Wittfeld[32,39], Kathryn Cullen[13], Liesbeth Reneman[7], Mardien L. Oudega[6], Margaret J. Wright[40], Maria J. Portella[41], Matthew D. Sacchet[42], Meng Li[11], Moji Aghajani[6,43], Mon-Ju Wu[9,10], Natalia Jaworska[44], Neda Jahanshad[15], Nic J. A. van der Wee[45], Nynke Groenewold[3], Paul J. Hamilton[46,47], Philipp G. Sämann[48], Robin Bülow[49], Sara Poletti[29], Sarah Whittle[50], Sophia I. Thomopoulos[15], Steven J.A. van, der Werff[51], Sheri-Michelle Koopowitz[3], Thomas Lancaster[21,22], Tiffany C. Ho[52,53], Tony T. Yang[52], Zeynep Basgoze[13], Dick J. Veltman[6], Lianne Schmaal[28], Paul M. Thompson[15], and Roberto Goya-Maldonado[1,*], for the ENIGMA Major Depressive Disorder working group[54]

Affiliations:

[1] Laboratory of Systems Neuroscience and Imaging in Psychiatry (SNIP-Lab), Department of Psychiatry and Psychotherapy, University Medical Center Goettingen (UMG), Georg-August University, Von-Siebold-Str. 5, 37075 Goettingen, Germany;

[2] SoCAT Lab, Department of Psychiatry, School of Medicine, Ege University, Izmir, Turkey;

[3] Department of Psychiatry & Mental Health, University of Cape Town, Cape Town, South Africa;

[4] Center for Neuroscience, University of Pittsburgh, Pittsburgh, PA, USA; Center for Neural Basis of Cognition, University of Pittsburgh, Pittsburgh, PA, USA;

[5] Department of Biomedical Sciences of Cells and Systems, University Medical Center Groningen, University of Groningen, Groningen, the Netherlands;

[6] Department of Psychiatry, Amsterdam UMC, Vrije Universiteit Amsterdam, Amsterdam Neuroscience, Amsterdam Public Health Research Institute, Amsterdam, The Netherlands;

[7] Amsterdam University Medical Centers, location AMC, Department of Radiology and Nuclear Medicine, Amsterdam, the Netherlands;

[8] Melbourne Neuropsychiatry Centre, Department of Psychiatry, The University of Melbourne, Parkville, Victoria, Australia;

[9] Louis A. Faillace, MD, Department of Psychiatry and Behavioral Sciences, The University of Texas Health Science Center at Houston, Houston, TX, USA;

[10] Center Of Excellence On Mood Disorders, Louis A. Faillace, MD, Department of Psychiatry and Behavioral Sciences at McGovern Medical School, The University of Texas Health Science Center at Houston, TX, USA;

[11] Department of Psychiatry and Psychotherapy, Jena University Hospital, Jena, Germany;

[12] Department of Psychology, University of Minnesota, Minneapolis, MN, USA;

[13] Department of Psychiatry and Behavioral Science, University of Minnesota Medical School, Minneapolis, MN, USA;

[14] Section on the Neurobiology and Treatment of Mood Disorders, National Institute of Mental Health, Bethesda, MD, USA;

[15] Imaging Genetics Center, Mark & Mary Stevens Neuroimaging and Informatics Institute, Keck School of Medicine, University of Southern California, Marina del Rey, CA 90274, USA;

[16] Department of Biomedical Sciences, Florida Stateu University, Tallahassee FL, USA;

[17] School of Psychology, University of East London, London, UK;

[18] Centre for Affective Disorders, Institute of Psychiatry, Psychology and Neuroscience, King's College London, London, UK;

[19] SA MRC Research Unit on Risk & Resilience in Mental Disorders, Department of Psychiatry & Neuroscience Institute, University of Cape Town, Cape Town, South Africa;

[20] Department of Psychology, School of Arts and Social Sciences, City, University of London, London, UK;

[21] Department of Neuroimaging, Institute of Psychiatry, Psychology and Neuroscience, King's College London, London, UK

[22] Cardiff University Brain Research Imaging Center, Cardiff University, Cardiff, UK;

[23] MRC Center for Neuropsychiatric Genetics and Genomics, Cardiff University, Cardiff, UK;

[24] Division of Psychological Medicine and Clinical Neurosciences, Cardiff University, Cardiff, UK;

[25] School of Mental Health and Neuroscience, Faculty of Health, Medicine and Life Sciences, Maastricht University, Maastricht, 6229 ER, The Netherlands;

[26] Department of Psychiatry, Psychotherapy and Psychosomatics, Medical School, RWTH Aachen University, Germany;

[27] FIDMAG Germanes Hospitalàries Research Foundation, Centro de Investigación Biomédica en Red de Salud Mental (CIBERSAM), Barcelona, Catalonia, Spain

[28] Orygen, The National Centre of Excellence in Youth Mental Health, Parkville, VIC, Australia;

[29] Centre for Youth Mental Health, The University of Melbourne, Parkville, VIC, Australia;

[30] Division of Neuroscience, IRCCS San Raffaele Scientific Institute, Milano, Italy;

[31] Departments of Psychiatry and Pediatrics, University of Calgary, Calgary, AB, Canada;

[32] Department of Psychiatry and Psychotherapy, University Medicine Greifswald, Greifswald, Germany;

[33] Institute for Community Medicine, University Medicine Greifswald, Greifswald, Germany;

[34] Department of Psychology, Stanford University, Stanford, CA, USA;

[35] Experimental Therapeutics and Pathophysiology Branch, National Institute for Mental Health, National Institutes of Health, Bethesda, MD, USA;

[36] West Region, Institute of Mental Health, Singapore;

[37] Yong Loo Lin School of Medicine, National University of Singapore, Singapore;

[38] Lee Kong Chian School of Medicine, Nanyang Technological University, Singapore;

[39] German Center for Neurodegenerative Diseases (DZNE), Site Rostock/ Greifswald, Greifswald, Germany;

[40] Queensland Brain Institute, The University of Queensland, Brisbane, QLD, Australia;

[41] Sant Pau Mental Health Research Group, Institut de Recerca de l'Hospital de la Santa Creu i Sant Pau, Barcelona, Catalonia, Spain. CIBERSAM, Madrid, Spain;

[42] Meditation Research Program, Department of Psychiatry, Massachusetts General Hospital, Harvard Medical School, Boston, MA, USA;

[43] Institute of Education & Child Studies, Section Forensic Family & Youth Care, Leiden University, Leiden, The Netherlands;

[44] Department of Psychiatry, McGill University, Montreal, Quebec, Canada;

[45] Deparment of Psychiatry, Leiden Institute for Brain and Cognition and Theme NeuroscienceLeiden University Medical Center, Netherlands;

[46] Center for Social and Affective Neuroscience, Department of Biomedical and Clinical Sciences, Linköping University, Linköping, Sweden;

[47] Center for Medical Imaging and Visualization, Linköping University, Linköping, Sweden;

[48] Max Planck Institute of Psychiatry, Munich, Germany;

[49] Institute for Radiology and Neuroradiology, University Medicine Greifswald, Greifswald, Germany;

[50]Melbourne Neuropsychiatry Centre, Department of Psychiatry, The University of Melbourne & Melbourne Health, Melbourne, VIC, Australia;

[51] Department of Psychiatry, Leiden University Medical Center, Leiden, Netherland;

[52] Department of Psychiatry and Behavioral Sciences, Division of Child and Adolescent Psychiatry, Weill Institute for Neurosciences, University of California, San Francisco, San Francisco, CA, USA;

[53] Department of Psychology, University of California, Los Angeles, Los Angeles, CA, USA;

[54] https://enigma.ini.usc.edu/ongoing/enigma-mdd-working-group/

***Corresponding author:**
PD Dr. Roberto Goya-Maldonado
Laboratory of Systems Neuroscience and Imaging in Psychiatry (SNIP-Lab)
Department of Psychiatry and Psychotherapy
University Medical Center Göttingen (UMG)
Von-Siebold Str. 5, 37075 Göttingen
e-mail: roberto.goya@med.uni-goettingen.de1

## Abstract

Machine learning (ML) techniques have gained popularity in the neuroimaging field due to their potential for classifying neuropsychiatric disorders. However, the diagnostic predictive power of the existing algorithms has been limited by small sample sizes, lack of representativeness, data leakage, and/or overfitting. Here, we overcome these limitations with the largest multi-site sample size to date (n=5,356) to provide a generalizable ML classification benchmark of major depressive disorder (MDD). Using brain measures from standardized ENIGMA analysis pipelines in FreeSurfer, we were able to classify MDD vs healthy controls (HC) with around 62% balanced accuracy, but when harmonizing the data using ComBat balanced accuracy dropped to approximately 52%. Similar results were observed in stratified groups according to age of onset, antidepressant use, number of episodes and sex. Future studies incorporating higher dimensional brain imaging/phenotype features, and/or using more advanced machine and deep learning methods may achieve more encouraging prospects.

## Introduction

Major depressive disorder (MDD) is a psychiatric disorder with great impact on society, with a lifetime prevalence of 14% [1], often resulting in reduced quality of life [2] and increased risk of suicide for those affected [3]. Considering the possibility of treatment resistance [4] and accelerated brain aging [5], early recognition and implementation of effective treatments are critical. Unfortunately, there are no reliable biomarkers to date to diagnose MDD, to predict its highly variable natural progression or response to treatment [6]. Until now, the diagnosis of MDD relies exclusively on self-reported symptoms in clinical interviews, which - despite great efforts - present risk of misdiagnosis due to subjectivity and limited specificity of some symptoms, especially in the early stage of mental disorders. Furthermore, comorbid conditions such as substance use disorders, anxiety spectrum disorders [7], and other mental and somatic diseases [8] may contribute to the difficulty of correctly diagnosing and treating MDD.

With modern neuroimaging techniques such as magnetic resonance imaging (MRI), it has become possible to investigate cortical and subcortical brain alterations associated with MDD with high spatial resolution. Numerous studies reveal structural brain differences in MDD compared to healthy controls (HC) [9–13], with patients presenting, on average, smaller hippocampal volumes as well as lower cortical thickness in the insula, temporal lobes, and orbitofrontal areas. However, inference at the group level and small effect sizes preclude clinical application. Analytic tools such as machine learning (ML) that allow multivariate combinations of brain features and enable inference at the individual level may result in better discrimination between MDD 245patients and HC, thereby potentially providing clinically relevant biomarkers for MDD.

Current literature shows MRI-based MDD classification accuracies ranging from 53% to 91% [14,15] with inconsistencies regarding which regions are the most informative for the classification. This lack of consensus in the literature raises concerns regarding the generalizability of the classification methods and their related findings. A major contributor to high variability in classification performances is sample size [16,17]. Specifically smaller samples of the test data set tend to show more extreme results in both directions [16], whereas studies with larger sample sizes in the test set tend to converge to an accuracy of around 60% [17]. In the presence of publication bias, which favors the reporting of overestimations, published literature can quickly accumulate inflated results [18]. Further, overestimations in the neuroimaging field [19,20] may also be driven by data leakage, which refers to the use of test data in any part of the training process.

Other factors contributing to inconsistencies in results include the heterogeneity of samples in relation to demographic and clinical characteristics, factors that play a significant role both in MDD and in the general population [5,21,22]. As large representative samples within a single cohort is difficult (e.g., due to financial cost, access to patient population, etc.), there is a growing interest in performing multi-site mega-analyses to address these issues.

ENIGMA MDD is a large-scale worldwide consortium, which curates and applies standardized analysis protocols to MRI and clinical/demographic data of MDD patients and HC from 52 independent sites from 17 countries across 6 continents (for review, [23]). Such large-scale approaches with global representation are necessary for identifying brain alterations associated with MDD that are realistic, reliable, and generalizable [24]. Therefore, we consider data from different international cohorts included in ENIGMA MDD a powerful and efficient resource to benchmark the robustness of the most commonly used ML algorithms in MDD-related neuroimaging studies (for review, [14]). Such algorithms include support vector machines (SVM), logistic regression with least absolute shrinkage and selection operator (LASSO) and ridge regularization, elastic net, and random forests. An additional advantage of ENIGMA MDD is that the inclusion of thousands of participants allows the stratification of several important factors related to cortical and subcortical brain alterations in MDD such as sex, age, age of MDD onset, and antidepressant use. However, unifiying multi-site data presents challenges. The global group differences between cohorts - referred to here as a site effect - may arise from different MR acquisition equipment and acquisition protocols [25], and/or demographic and clinical factors [26,27]. Ignoring the site effect may lead to construction of suboptimal less-generalizable classification models [28], hindering the generalizability of the results. Along these lines, a commonly used strategy to mitigate site effect is to apply a harmonization technique such as ComBat [29]. Adopted from genomic studies, NeuroComBat estimates and statistically corrects for (harmonizes) differences in location (mean) and scale (variance) across different cohorts, while preserving or perhaps even enhancing the effect size of the variables of interest [30–32]. There are only a few studies attempting multi-site MDD classification using structural brain metrics [17,33,34]; however, site effects were not addressed in their analyses.

The main goal of this study was to establish a benchmark for classification of MDD vs HC based on structural cortical and subcortical brain measures in the largest sample to date. We applied the most commonly used ML algorithms in MDD neuroimaging studies: SVM with linear and rbf kernels with and without feature selection (PCA, t-test), logistic regression with LASSO/ridge regularization, elastic net, and random forests. The model's performance is estimated via balanced accuracy, area under the receiver operating characteristic (AUC), sensitivity and specificity. We hypothesized that all models would be able to classify MDD vs HC with balanced accuracy higher than random chance, based on provided brain measures. We pooled preprocessed structural data from ENIGMA MDD participants, including 5,365 subjects (2,288 MDD and 3,077 HC) from 30 cohorts worldwide. As we were equally interested in general structural brain alterations in MDD as well as the generalizability of classification performance in sites unseen in the training phase, the data were split according to two strategies. First, age and sex (Splitting by Age/Sex) were evenly distributed across all cross-validation (CV) folds, where each fold is used as a test set once and the rest of folds is used as a training set iteratively. Second, the sites (Splitting by Site) were kept whole across CV folds, so the algorithms were trained and tested on different sets of cohorts, resulting in large between-sample heterogeneity of training and test sets, potentially resulting in lower classification performance [35], especially if large site effects are present. Because MDD is a highly heterogeneous diagnosis - and previous work from ENIGMA MDD [10,11] has identified distinct alterations in different clinical subgroups - we also stratified MDD based on sex, age of onset, antidepressant use, and number of depressive episodes to investigate whether classification accuracy could be improved when considering more homogenous subgroups. Additionally, we investigated which brain areas were most relevant to classification performance.

In summary, we expected that (1) All models would correctly classify MDD above chance level, (2) Splitting by Site would yield lower performance versus Splitting by Age/Sex, (3) Application of ComBat would improve classification performances for all models, and (4) Stratified analyses according to demographic and clinical characteristics would yield higher classification performance. We also explored the impact of other approaches to remove site effects (ComBat-GAM [36] and CovBat [37]) from structural brain measures prior to feeding these measures into the classification models.

## Material and methods

*Participant Sample*

A total of 5,365 participants, 2,288 patients with MDD and 3,077 healthy controls, from 30 cohorts participating in the ENIGMA MDD working group, were included in the analyses. Information on sample characteristics, inclusion/exclusion criteria for each cohort can be found in Supplementary Table 1. Subjects with less than 75% of combined cortical and subcortical features and/or missing demographic/clinical information required for a particular analysis were excluded from the analysis. Missing cortical and subcortical features for the remaining subjects (2% of all data) were imputed by using multiple linear regression with age and sex of all subjects (regardless of diagnosis) as predictors, estimated for each cohort separately. All participating sites reported approval from their institutional review boards and local ethics committees and also obtained written informed consent for all participants.

*Brain Imaging Processing*

Structural T1-weighted 3D brain MRI scans of participating subjects were acquired from each site and preprocessed according to the rigorously validated ENIGMA Consortium protocols (http://enigma.ini.usc.edu/protocols/imaging-protocols/). Information on the MRI scanners and acquisition protocols used for each cohort can be found in Supplementary Table 2. To facilitate the ability to pool the data from different cohorts, cortical and subcortical parcellation was performed on every subject via the freely available FreeSurfer (Version 5.1, 5.3, 6 and 7.2) software [67,68]. Every cortical and subcortical brain parcellation was visually inspected as part of a careful quality check (QC) and statistically evaluated for outliers, according to the ENIGMA Consortium protocol (https://enigma.ini.usc.edu/protocols/imaging-protocols/). Cortical gray matter segmentation was based on the Desikan–Killiany atlas [69], yielding cortical surface area and cortical thickness measures for 68 brain regions (34 for each hemisphere), resulting in 136 cortical features. Subcortical segmentation was based on the *Aseg* atlas [69], providing volumes of 40 regions (20 for each hemisphere), of which we included 16: lateral ventricle, thalamus, caudate, putamen, pallidum, hippocampus, amygdala, and nucleus accumbens, bilaterally.

*Data Splitting into Cross-Validation Folds*

We applied two different strategies to split the data into training and test sets: *Splitting by Age/Sex* and *Splitting by Site*. For both strategies, data was split into 10 folds, 9 of which were used for the training and the remaining fold was used as a test set. This was repeated iteratively until each fold was used once as a test set, thus performing the 10-fold CV. We investigated the general differences in brain volumes that can characterize MDD by using the Splitting by Age/Sex strategy. In this way, the age and sex distribution as well as number of subjects between the folds were balanced to mitigate the effect of these factors on the classification performance. However, it should be noted that with each site represented in every CV fold the potential site effects in this strategy, if any, would be diluted between the folds, which would not represent a realistic clinical scenario, where a classification model likely has to generalize to unseen sites. Therefore, we used a second strategy, Splitting by Site, which would yield more realistic metrics of classification performance for unseen sites. Using this strategy, every site was present only in one fold, thus the model is always trained and tested on different sets of sites and sites were distributed across folds to balance the number of subjects in every fold as close as possible. In this scenario, potential site-specific confounders (e.g., different MR scanners/acquisition protocols, demographic and clinical differences, etc.) were not equally distributed between the training and test sets. In this way, we can fairly evaluate the generalizability from one cohort to another. Finally, to access the performance estimates for each site, we explored leave-site-out CVs.

*Analysis Pipeline*

After distributing the data into CV folds corresponding to the splitting strategies, 9 folds were used for the training, while the remaining fold was held out as a test set (Figure 3). CV folds were residualized normatively, partialling out the linear effect of age, sex and ICV from all cortical and subcortical features. In this step, age, sex and ICV regressors were estimated on the HC from training CV folds and applied to the MDD training data and to all data in the test set. After normalizing all features to have mean of zero and standard deviation of one based on the estimated from the training set's initial features' distributions, training and test folds were used for training and performance estimation, respectively. Additionally, class weighting was performed to mitigate an unbalanced training set across classes. Models' hyperparameters were estimated in the training data via nested 10-folds cross-validation using grid search (random splits, for both Splitting by Site and Splitting by Age/Sex), before the performance was measured on the test data to avoid data leakage through the choice of hyperparameters. The list of hyperparameters that were adjusted can be found in Supplementary Table 3. We evaluated the performance of SVM with linear kernel, SVM with rbf kernel, logistic regression with LASSO regularization, logistic regression with ridge regularization, elastic net, and random forests by using balanced accuracy, sensitivity, specificity and AUC as performance metrics. As regularization serves as an in-build feature selection algorithm, we additionally applied feature selection via PCA and t-test for SVM algorithms. To access model-level assessment [70], all models were also trained on the subset of features, i.e. only cortical surface areas, only cortical thicknesses and only subcortical volumes. Lastly, we investigated which features contributed most to the classification performance by looking at the decision-making of the most successful model. In case no performance differences across models were found, we reported the weights of the SVM with linear kernel as the representative classifier. These weights correspond to the classification performance of Splitting by Age/Sex strategy as all sites are used for weight's estimation. To assess confidence intervals of the feature weights, we performed 599-bootstrap [71] on the whole data set.

Further analyses were performed by stratifying the data according to demographic and clinical categories, including sex, age of onset (<21 years old vs >21 years old), antidepressant use (yes/no at time of scan), and number of depressive episodes (first episode vs recurrent episodes). The subjects with missing information on these factors were not included in these stratification analyses, while they were still considered for the main analysis.

All the steps from CV folds to classification were repeated with feature specific harmonization of site effects via ComBat. Variance explained by age, sex and ICV was preserved in the cortical and subcortical features during harmonization step. The harmonized folds were then residualized normatively with all subsequent steps from the analysis without harmonization step. Furthermore, we compared ComBat with two modifications: ComBat-GAM and CovBat. More detailed description of ComBat, ComBat-GAM and CovBat as well as their implementation for both splitting strategies can be found in Supplementary section "Harmonization methods".

We used Python (version 3.8.8) to perform all calculations. All classification models and feature selection methods were imported from sklearn library (version 1.1.2, https://scikit-learn.org/stable/). We modified ComBat script (https://github.com/Jfortin1/ComBatHarmonization) to incorporate ComBat-GAM (https://github.com/rpomponio/neuroHarmonize) and CovBat (https://github.com/andy1764/CovBat_Harmonization) in one function for both splitting strategies.

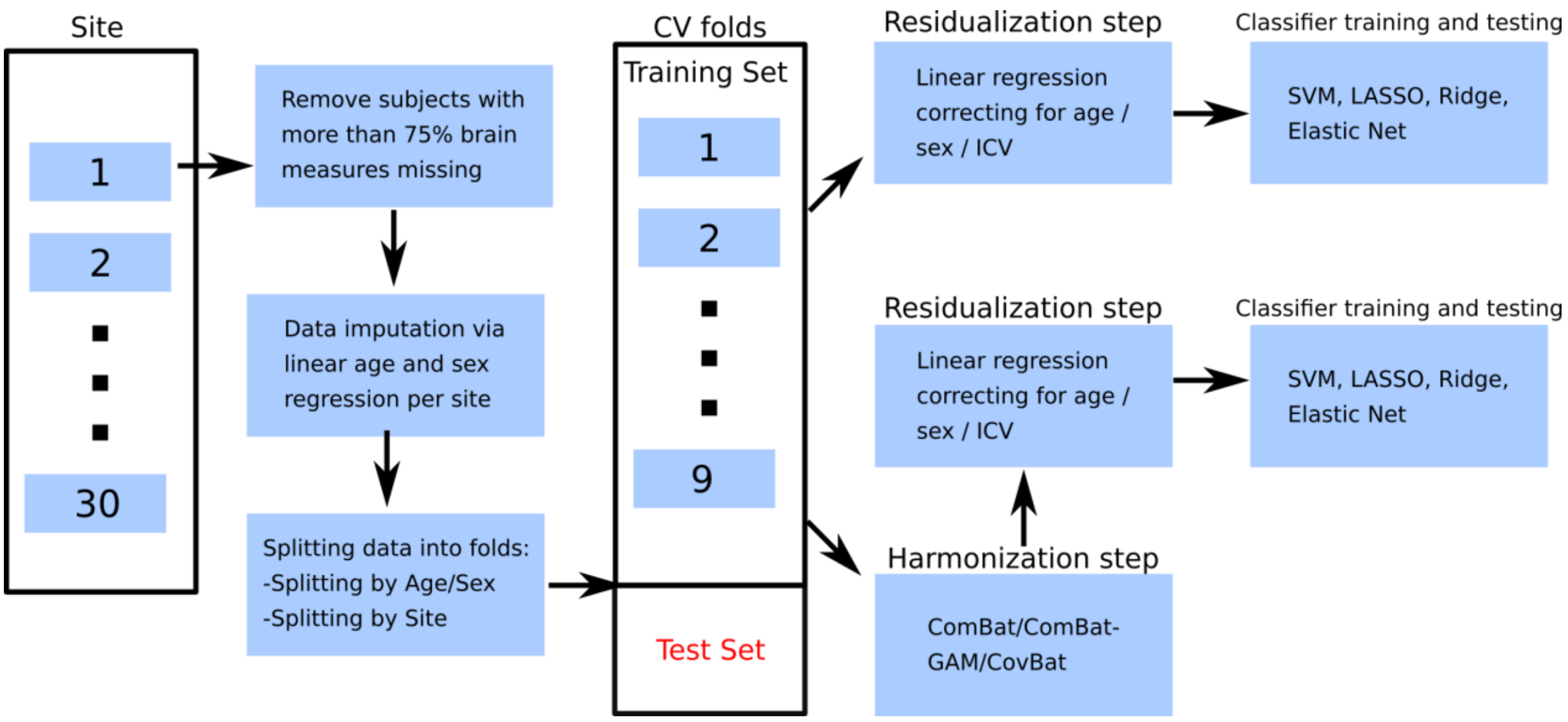

**Figure 3: Detailed analysis pipeline.** Initial data from all cohorts is split into training and test sets according to splitting strategies (Splitting by Age/Sex and Splitting by Site) after removing subjects with more than 75% missing data and data imputation steps. The corresponding training folds are then residualized directly to remove ICV, age and sex related effects and fed to the classification algorithms. In case of harmonization by ComBat, the residualization step takes place after the harmonization step is conducted. If training folds were harmonized by ComBat, the test fold was harmonized as well by using ComBat estimates from the training folds. Next, the test fold was residualized by using estimates obtained from the training folds. We estimated classification performance on the residualized test fold. This routine was performed iteratively for each combination of training and test folds.

## Results

*Participants and Data Splitting*

From 5,572 participants, 207 were excluded due to less than 75% of combined cortical and subcortical features being provided, resulting in 5,365 subjects (2,288 MDD and 3,077 HC) used in the analysis.

Substantial differences in age (87% of pairwise comparisons between cohorts were significant, t-test, $p<0.05$) and sex (54%, t-test, $p<0.05$) distribution exist in the investigated cohorts (Table 1, Supplementary Table 4). In the Splitting by Age/Sex strategy, all cohorts were evenly distributed across the folds, resulting in a similar number of subjects in each of fold (Table 2 left). In the Splitting by Site strategy, entire cohorts were kept into single folds, this time balancing the total number of subjects in each fold as close as possible (Table 2 right). This resulted in an irregular number of participants in each fold, with some folds containing only one of the larger cohorts (e.g., SHIP-T0, SHIP-S2, MPIP) and others containing multiple smaller cohorts.

**Table 1:** ENIGMA MDD participating cohorts in the study. Each cohort is presented with number of total subjects, number of patients with major depressive disorder (MDD) and healthy controls (HC), as well as their mean age (in years) and sex (number and % of females).

| Cohort | | Number of subjects | Age Mean (SD) | Number of Females (%) |
|---|---|---|---|---|
| **AFFDIS** | **Total** | 79 | 39.75 (14.67) | 36(46) |
| | **HC** | 46 | 39.87 (14.29) | 22(48) |
| | **MDD** | 33 | 39.58 (15.18) | 14(42) |
| **Pharmo (AMC)** | **Total** | 51 | 29.37 (4.64) | 51(100) |
| | **HC** | 0 | nan | nan |
| | **MDD** | 51 | 29.37 (4.64) | 51(100) |
| **Barcelona-StPau** | **Total** | 94 | 46.66 (7.81) | 72(77) |
| | **HC** | 32 | 46.03 (8.00) | 23(72) |
| | **MDD** | 62 | 46.98 (7.68) | 49(79) |
| **CARDIFF** | **Total** | 40 | 46.55 (11.74) | 27(68) |
| | **HC** | 0 | nan | nan |
| | **MDD** | 40 | 46.55 (11.74) | 27(68) |
| **CSAN** | **Total** | 109 | 34.70 (12.88) | 74(68) |
| | **HC** | 49 | 33.20 (12.07) | 34(69) |
| | **MDD** | 60 | 35.92 (13.38) | 40(67) |
| **Calgary** | **Total** | 107 | 17.03 (4.12) | 60(56) |
| | **HC** | 52 | 15.81 (5.03) | 29(56) |
| | MDD | 55 | 18.19 (2.51) | 31(56) |
| DCHS | Total | 79 | 30.91 (6.71) | 79(100) |
| | HC | 61 | 31.49 (6.82) | 61(100) |
| | MDD | 18 | 28.94 (5.89) | 18(100) |
| ETPB | Total | 60 | 35.03 (9.86) | 36(60) |
| | HC | 26 | 33.88 (10.22) | 16 (62) |
| | MDD | 34 | 35.91 (9.48) | 20(59) |
| Episca (Leiden) | Total | 49 | 15.00 (1.54) | 42(86) |
| | HC | 30 | 14.73 (1.53) | 26(87) |
| | MDD | 19 | 15.42(1.46) | 16(84) |
| FIDMAG | Total | 69 | 47.22 (12.29) | 44(64) |
| | HC | 34 | 45.94 (11.49) | 22(65) |
| | MDD | 35 | 48.46 (12.90) | 22(63) |
| Groningen | Total | 41 | 44.27 (13.67) | 30(73) |
| | HC | 21 | 44.05 (13.96) | 16(76) |
| | MDD | 20 | 44.50 (13.34) | 14(70) |
| Houston | Total | 290 | 28.72 (16.30) | 169(58) |
| | HC | 186 | 26.76 (15.91) | 105(56) |
| | MDD | 104 | 32.23 (16.39) | 64(62) |
| Jena | Total | 107 | 46.76 (15.00) | 52(49) |
| | HC | 77 | 47.75 (15.93) | 36(47) |
| | MDD | 30 | 44.20 (11.92) | 16(53) |
| LOND | Total | 130 | 49.67 (8.62) | 79(61) |
| | HC | 61 | 51.72(7.87) | 32(53) |
| | MDD | 69 | 47.86(8.85) | 47(68) |
| MODECT | Total | 42 | 72.71 (9.25) | 28(67) |
| | HC | 0 | nan | nan |
| | MDD | 42 | 72.71 (9.25) | 28(67) |
| MPIP | Total | 548 | 48.66 (13.59) | 315(57) |
| | HC | 211 | 49.53 (13.02) | 124 (59) |
| | MDD | 337 | 48.12 (13.90) | 191(57) |
| Melbourne | Total | 245 | 19.42 (2.88) | 130(53) |
| | HC | 102 | 19.58 (2.97) | 54(53) |
| | MDD | 143 | 13.31 (2.80) | 76(53) |
| Minnesota | Total | 110 | 15.47 (1.89) | 79(72) |
| | HC | 40 | 15.68 (1.98) | 26(65) |
| | MDD | 70 | 15.36 (1.83) | 53(76) |
| Moraldilemma | Total | 70 | 18.81 (1.94) | 70(100) |
| | HC | 46 | 18.50 (1.75) | 46(100) |
| | MDD | 24 | 19.42 (2.14) | 24(100) |
| NESDA | Total | 219 | 38.11 (10.32) | 145(66) |
| | HC | 65 | 40.29 (9.67) | 42(65) |
| | MDD | 154 | 37.19 (10.45) | 103(67) |
| QTIM | Total | 386 | 22.08 (3.25) | 267(69) |
| | HC | 284 | 22.11 (3.30) | 190(67) |
| | MDD | 102 | 22.01 (3.11) | 77(75) |
| UCSF | Total | 163 | 15.46 (1.31) | 91(56) |
| | HC | 88 | 15.32 (1.28) | 42(48) |
| | MDD | 75 | 15.63 (1.33) | 49(65) |
| SHIP_S2 | Total | 579 | 55.01 (12.57) | 294(51) |
| | HC | 443 | 55.44 (12.80) | 198(45) |
| | MDD | 136 | 53.59 (11.68) | 96(71) |
| SHIP_T0 | Total | 1229 | 50.15 (13.69) | 607(49) |
| | HC | 919 | 50.50 (14.18) | 405(44) |
| | MDD | 310 | 49.12 (12.04) | 202 (65) |
| SanRaffaele | Total | 45 | 49.07 (13.51) | 32(71) |
| | HC | 0 | nan | nan |
| | MDD | 45 | 49.07 (13.51) | 32(71) |
| Singapore | Total | 38 | 39.50 (6.43) | 18(47) |
| | HC | 16 | 38.69 (4.59) | 8(50) |
| | MDD | 22 | 40.09 (7.43) | 10(45) |
| Socat_dep | Total | 179 | 37.85 (13.34) | 161(90) |
| | HC | 100 | 36.42 (13.57) | 90 (90) |
| | MDD | 79 | 39.66 (12.81) | 71 (90) |
| StanfFAA | Total | 32 | 32.71 (9.56) | 32(100) |
| | HC | 18 | 30.44 (9.96) | 18(100) |
| | MDD | 14 | 35.63 (8.14) | 14(100) |
| StanfT1wAggr | Total | 115 | 37.18 (10.27) | 69(60) |
| | HC | 59 | 37.24 (10.43) | 36(61) |
| | MDD | 56 | 37.11 (10.09) | 33(59) |
| TIGER | Total<br>HC<br>MDD | 60<br>11<br>49 | 15.63 (1.34)<br>15.18 (1.03)<br>15.73 (1.38) | 38(63)<br>5(45)<br>33(67) |
| All sites | Total<br>HC<br>MDD | 5365<br>3077<br>2288 | 39.84 (17.69)<br>40.82(18.09)<br>38.52 (17.05) | 3227(60)<br>1706(55)<br>1521(66) |

**Table 2: Data splitting strategies.** The differences in strategies are seen in the distribution of age, sex, and diagnosis between cross-validation folds.

| Splitting By Age/Sex | | | | Splitting By Site | | | |
|---|---|---|---|---|---|---|---|
| Fold | Age mean (SD) | Number of Females (%) | Number of subjects (%MDD) | Fold | Age mean (SD) | Number of Females (%) | Number of subjects (%MDD) |
| 1 | 39.98 (17.40) | 322 (60) | 536 (42) | 1 | 50.15 (13.69) | 607 (49) | 1229 (25) |
| 2 | 39.63 (17.81) | 324 (60) | 538 (42) | 2 | 55.01 (12.57) | 294 (51) | 579 (23) |
| 3 | 39.85 (17.57) | 325 (60) | 538 (43) | 3 | 48.66 (13.59) | 315 (57) | 548 (61) |
| 4 | 39.66 (17.94) | 322 (60) | 535 (39) | 4 | 22.90 (4.97) | 299 (72) | 418 (28) |
| 5 | 39.99 (17.56) | 323 (60) | 538 (44) | 5 | 36.72 (19.69) | 272 (60) | 451 (51) |
| 6 | 39.75 (17.25) | 317 (60) | 531 (43) | 6 | 22.53 (10.92) | 293 (65) | 450 (68) |
| 7 | 40.15 (17.89) | 327 (60) | 541 (42) | 7 | 35.94 (12.96) | 295 (71) | 418 (59) |
| 8 | 39.81 (17.93) | 322 (60) | 535 (44) | 8 | 38.85 (12.66) | 348 (81) | 431 (45) |
| 9 | 39.86 (17.73) | 320 (60) | 535 (44) | 9 | 24.79 (16.16) | 203 (54) | 377 (42) |
| 10 | 39.74 (17.80) | 325 (60) | 538 (43) | 10 | 34.95 (15.45) | 301 (65) | 464 (55) |

*Full Data Set Analysis*

The classification performance of all models was similar and is presented in Table 3. When sites were evenly distributed across all CV folds (Splitting by Age/Sex), the highest balanced accuracy of 0.639 was achieved by SVM with rbf kernel, when trained using all cortical and subcortical features. The application of ComBat harmonization resulted in a performance drop of all models close to chance level. This pattern of lower classification performance, when ComBat was applied, was also observed across other classification metrics (see Supplementary Table 5-7). Yet specificity was found to be up to 10% higher than sensitivity, possibly related to potential imbalances in ratio MDD to HC and its effect on the classification. For the Splitting by Site strategy, classification performances did not change significantly based on whether the ComBat harmonization was performed or not. Balanced accuracy was close to random chance, indicating that the models were not able to differentiate MDD subjects from HC. The application of ComBat did not result in higher classification accuracies (Table 3). By exploring the classification performances measured on only a subset of cortical and subcortical features, we observed very similar results with classification around chance level. Similarly, there was no improvement when more sophisticated harmonization algorithms such as ComBat-GAM and CovBat were applied (see Supplementary Table 8).

**Table 3:** Balanced accuracy measured on the entire data set, after being divided into cross-validation folds using the Splitting by Age/Sex and Splitting by Site strategies. We evaluated classification performances when models are trained on combined cortical and subcortical features , cortical thickness, cortical surface area, and subcortical volume. Furthermore, all models were trained/tested without and with ComBat harmonization.

| **Splitting by Age/Sex** | | | | | | | | |
|---|---|---|---|---|---|---|---|---|
| | Cortical + Subcortical | | Cortical Thickness | | Cortical Surface area | | Subcortical Volume | |
| | No ComBat | With ComBat | No ComBat | With ComBat | No ComBat | With ComBat | No ComBat | With ComBat |
| **SVM linear** | 0.616 | 0.524 | 0.577 | 0.504 | 0.572 | 0.518 | 0.602 | 0.524 |
| **SVM rbf** | 0.639 | 0.525 | 0.600 | 0.515 | 0.578 | 0.510 | 0.619 | 0.513 |
| **SVM + PCA** | 0.638 | 0.529 | 0.601 | 0.513 | 0.575 | 0.518 | 0.622 | 0.513 |
| **SVM + ttest** | 0.627 | 0.515 | 0.581 | 0.515 | 0.567 | 0.526 | 0.619 | 0.521 |
| **LASSO** | 0.612 | 0.524 | 0.583 | 0.499 | 0.578 | 0.516 | 0.596 | 0.518 |
| **Ridge** | 0.610 | 0.523 | 0.585 | 0.498 | 0.573 | 0.515 | 0.594 | 0.520 |
| **Elastic Net** | 0.609 | 0.523 | 0.584 | 0.500 | 0.569 | 0.517 | 0.593 | 0.520 |
| **Random Forests** | 0.613 | 0.507 | 0.593 | 0.514 | 0.574 | 0.509 | 0.611 | 0.511 |
| **Splitting by Site** | | | | | | | | |
| | Cortical + Subcortical | | Cortical Thickness | | Cortical Surface area | | Subcortical Volume | |
| | No ComBat | With ComBat | No ComBat | With ComBat | No ComBat | With ComBat | No ComBat | With ComBat |
| **SVM linear** | 0.512 | 0.519 | 0.498 | 0.495 | 0.499 | 0.506 | 0.506 | 0.521 |
| **SVM rbf** | 0.513 | 0.515 | 0.493 | 0.499 | 0.493 | 0.513 | 0.503 | 0.519 |
| **SVM + PCA** | 0.527 | 0.520 | 0.502 | 0.512 | 0.504 | 0.524 | 0.520 | 0.520 |
| **SVM + ttest** | 0.502 | 0.512 | 0.487 | 0.499 | 0.507 | 0.508 | 0.510 | 0.527 |
| **LASSO** | 0.513 | 0.517 | 0.491 | 0.489 | 0.508 | 0.513 | 0.507 | 0.512 |
| **Ridge** | 0.514 | 0.514 | 0.497 | 0.490 | 0.505 | 0.509 | 0.507 | 0.514 |
| **Elastic Net** | 0.513 | 0.514 | 0.498 | 0.489 | 0.503 | 0.514 | 0.507 | 0.514 |
| **Random Forests** | 0.518 | 0.506 | 0.495 | 0.501 | 0.491 | 0.503 | 0.519 | 0.501 |

When no harmonization step was applied, the choice of CV splitting strategy affected all measures of classification performance. Splitting by Age/Sex strategy yielded a balanced accuracy above 0.60 compared to roughly 0.51 accuracy for the Splitting by Site strategy. The ComBat harmonization step evened the classification performance of algorithms for the different splitting strategies, both being close to random chance. Information on the balanced accuracy changes via ComBat performing leave-one-site-out CV can be found in Supplementary Table 9.

As the performance of the models were similar across all conditions, we accessed the weights of SVM with linear kernel to investigate, which regions contributed the most to the classification. The performance of SVM with and without application of ComBat was primarily driven by roughly the same set of cortical features, which could be observed by examining the feature weights. Feature weights of the SVM with linear kernel are presented in Figures 1 and 2. Even though the harmonization step affected the weights of the features, most of the informative features (with absolute weight >0.1) remained present. Cortical thickness features had greater weights compared to cortical surface areas, among which the left caudal middle frontal, left inferior parietal, left and right inferior temporal, left medial orbitofrontal, left postcentral, left precuneus, left superior frontal, right lingual, right paracentral, and right superior temporal regions were informative with and without the harmonization step. In the case of the regional surface areas, left and right cuneus, left inferior temporal, left medial orbitofrontal, left postcentral, and right precentral regions were found to be most informative for classification. Among subcortical volumes, no features remained informative after removing site effect via ComBat.

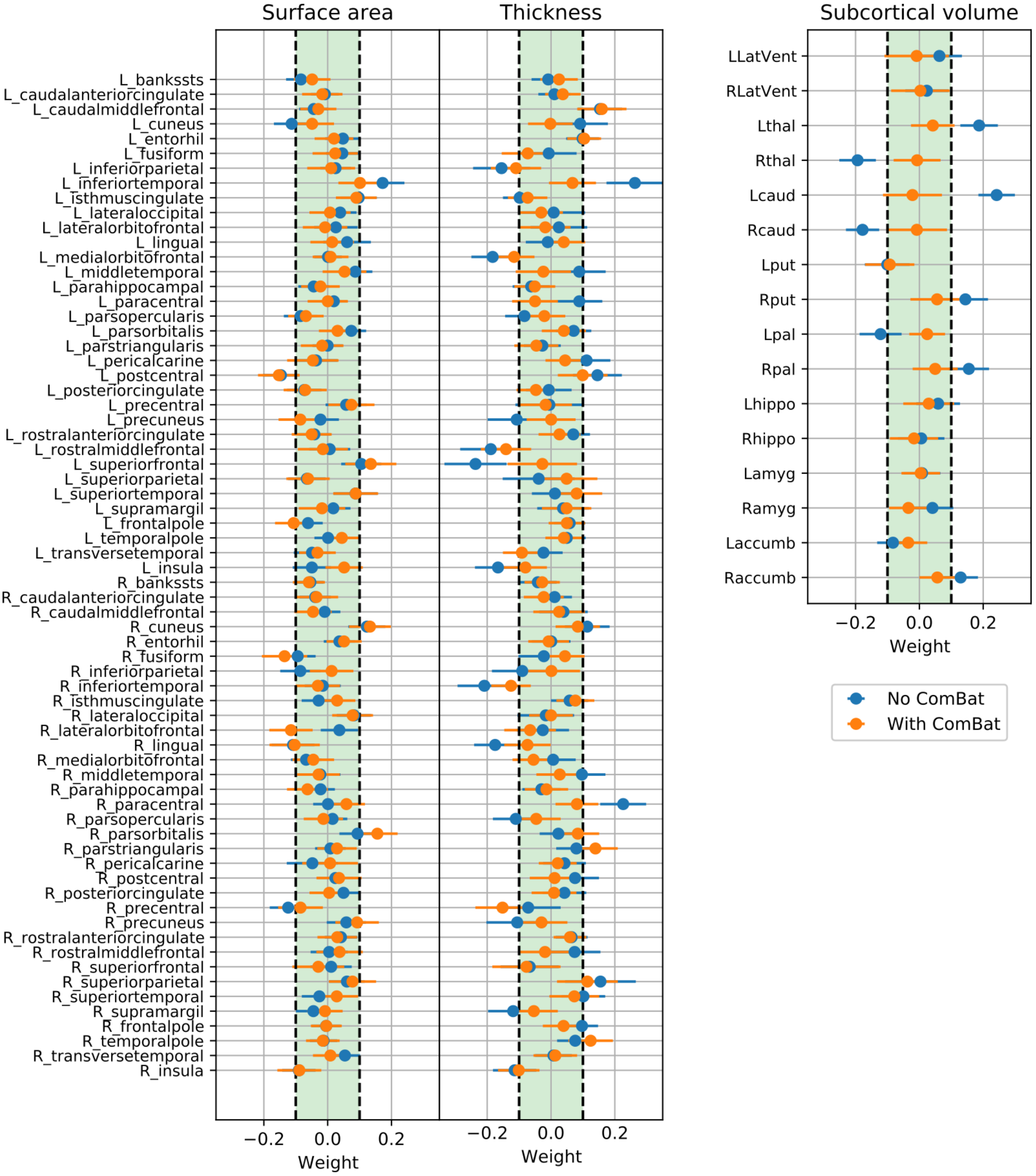

**Figure 1:** Feature weights of support vector machines (SVM) with the linear kernel. To assess the decision-making of SVM to differentiate subjects with major depressive disorder (MDD) from healthy controls (HC), we investigate the importance of the structural brain features by looking at the corresponding feature weights for the regional cortical surface areas, cortical thicknesses and subcortical volumes. The dashed lines represents ±0.1 as thresholds of feature weights, above which informative features stand out.

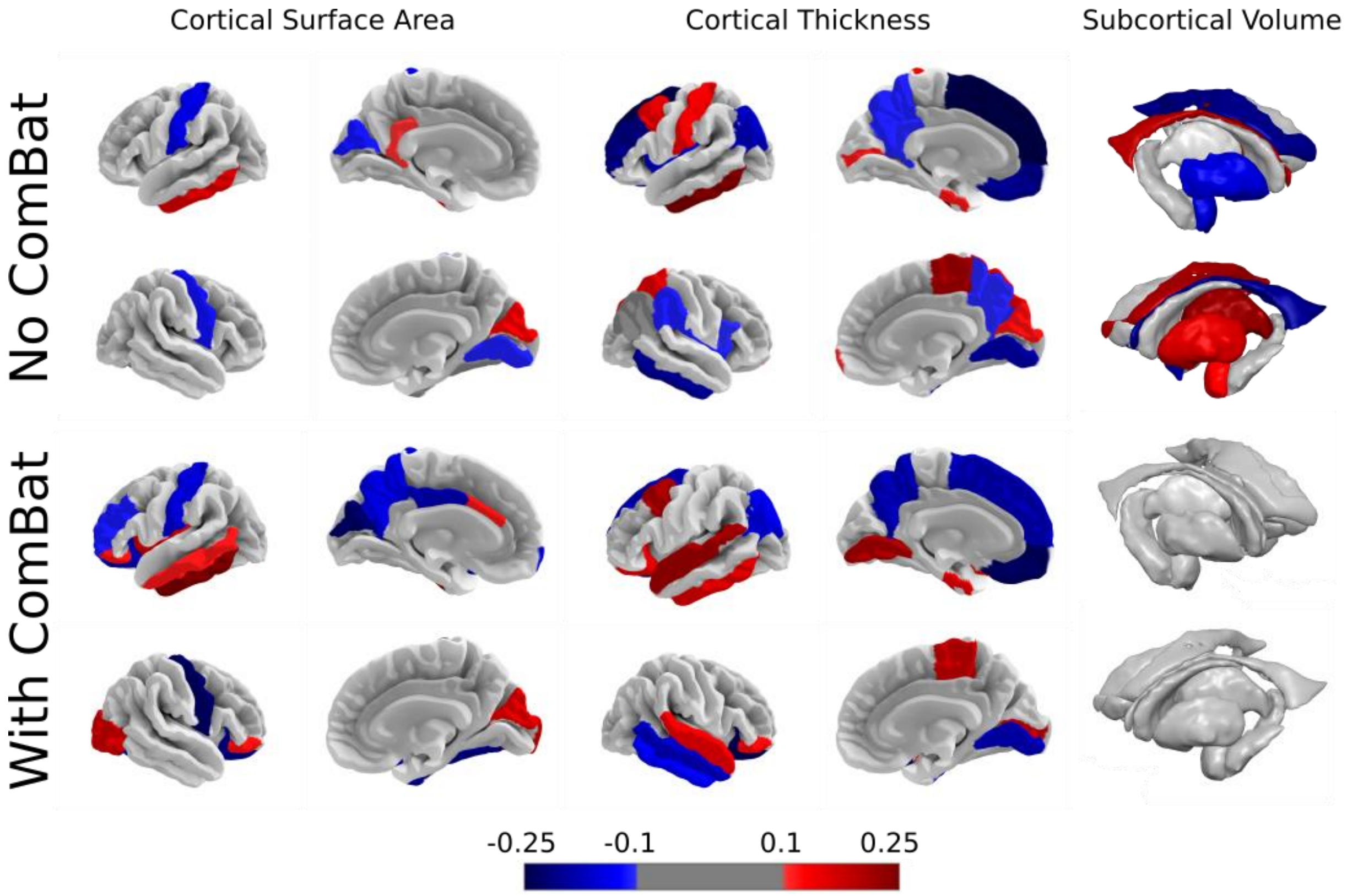

**Figure 2:** The most informative features for classification including regional cortical surface areas, thicknesses and subcortical volumes, trained on the whole data set without and with ComBat harmonization. Increased and decreased feature weight values in the major depressive disorder (MDD) group are represented by red and blue colormap, respectively.

*Data Stratification*

Next, we investigated the classification performance of models trained and tested on stratified data by demographic and clinical characteristics. The general pattern of the highest accuracy achieved by Splitting by Age/Sex strategy without ComBat and the significant drop in the accuracy when ComBat is applied was observed in all stratified analyses (below). In the Splitting by Site strategy, the classification performance did not change significantly when ComBat was applied. Information on the feature weights may be found in Supplementary Figures 1-4.

*Males vs females*

The number of male subjects is 2,131 and female subjects is 3,227 (7 male participants from the Episca cohort were not considered as we could not split them into 10 CV folds). In the Splitting by Age/Sex strategy without the harmonization step, the highest balanced accuracy of 0.632 was achieved when trained and tested on males - compared to maximum of 0.585 for females. When ComBat was applied, the accuracy dropped to 0.530 for males and to 0.529 for females, showing that there were minimal differences in classification results for males and females. For Splitting by Site, the accuracy did not change depending on the use of ComBat for both males (0.513 to 0.506) and females (0.519 to 0.517). Nevertheless, different brain regions were found important for classification in subgroups. In general, more features were found to be important for classification for males compared to females; this is especially noticeable for the regional surface areas (Supplementary Figure 1).

*Age of onset*

For Splitting by Age/Sex, when only patients first diagnosed in adolescence were included in the analysis, yielding 3,794 subjects in total, an accuracy of 0.626 was achieved, compared to 0.623 when patients who were first diagnosed in adulthood were analyzed. These accuracies dropped to 0.548 and 0.521 respectively, when ComBat was applied. In the Splitting by Site strategy, the balanced accuracy metrics did not change substantially for both subgroups: 0.541 to 0.544 for the adolescent-onset group and 0.546 to 0.518 for the adult-onset group, highlighting the absence of significant differences between these groups (Supplementary Figure 2).

*Antidepressant use vs antidepressant free (at the time of MR scan)*

Both subgroups showed a drop in balanced accuracy when ComBat was applied. In case of Splitting by Age/Sex, it reduced from 0.564 to 0.529 for the antidepressant-free subgroup (4,408 subjects) and from 0.716 to 0.534 for the antidepressant subgroup (3,988 subjects). When Splitting by Site, the balanced accuracy metrics did not change significantly for any of the subgroups when ComBat was used. For the antidepressant-free subgroup, it decreased from 0.564 to 0.528, while for the antidepressant group, it dropped from 0.560 to 0.483 (Supplementary Figure 3).

*First episode vs recurrent episodes*

Similarly, a drop in accuracy was observed when the data set was stratified based on the number of depressive episodes with vs without ComBat. In Splitting by Age/Sex, the balanced accuracy for the first episode subgroup dropped from 0.559 to 0.518 when ComBat was applied. For individuals with more than one episode, the balanced accuracy decreased from 0.644 to 0.520 with ComBat. In the Splitting by Site strategy, the algorithm's performance was not majorly affected by ComBat in the single episode subgroup, yielding 0.482 to 0.512 in balanced accuracy and an insignificant drop from 0.521 to 0.505 for the recurrent episodes subgroup (Supplementary Figure 4).

## Discussion

In this work, we benchmarked ML performance on the largest multi-site data set to date, using regional cortical and subcortical structural information for the task of discriminating patients with MDD vs HC. We applied the most commonly used ML algorithms to 152 features of 5,365 subjects from the ENIGMA MDD working group. To investigate brain characteristics common to MDD, as well as realistic classification metrics for unseen sites, we used two different data splitting approaches. Balanced accuracy was up to 63%, when data was split into folds according to *Splitting by Age/Sex*, and up to 51%, when data was split into folds according to *Splitting by Site* strategy. The harmonization of the data via ComBat evened the classification performance for both data splitting strategies, yielding up to 52% of balanced accuracy. This classification level implies that initial differences in performances were due to the site effects, most likely stemming from differences in MRI acquisition differences across sites. Lastly, the data set was stratified based on demographic and clinical factors, but we found only minor differences in terms of classification performances between subgroups.

*Data Splitting and Site Effect*

Splitting of the data plays an important role in formulating and testing the hypotheses as well as validating them. As shown in [38], different data splitting techniques in combination with machine and deep learning algorithms in medical mega-analytical studies may introduce unwanted biases influencing classification or regression performances. Here we aimed to consider two data splitting paradigms: Splitting by Age/Sex and Splitting by Site. With Splitting by Age/Sex, we investigated general MDD alterations in contrast to HC using ML methods to obtain unbiased results regarding these important demographic factors. When we look at the weights of the SVM with linear kernel estimated on the entire data set, they correspond to the performance from Splitting by Age/Sex, as every CV fold contains all sites and demographically corresponds closely to the whole data set. With Splitting by Site, we wanted to see if the knowledge learned in one subset of cohorts could be translated to unseen cohorts - this can only be realistically measured when data is split according to the site it belongs to. To the best of our knowledge, this is the first study to systematically emphasize differences in MDD vs HC classification performance in the context of data splitting strategies and the impact of ComBat in these strategies. The balanced accuracy of algorithms trained on data from Splitting by Age/Sex was up to 10% higher compared to Splitting by Site, confirming our expectations. This is a common trend in multi-site neuroimaging analyses [39], which indicates site effect and emphasizes how the nuances in data splitting strategies can strongly influence the classification performance. The presence of the site effect was additionally confirmed by training the SVM model to classify subjects according to their respective site, yielding substantially higher balanced accuracy compared to the main task of MDD vs HC classification (see Supplementary section "Harmonization methods"). The possibility that the site effect still reflected the demographic differences across cohorts, as cortical and subcortical features undergo substantial changes throughout lifespan [36] and differ between males and females [21,22], was not supported. Regressing out these sources of demographic information did not significantly change the level of classification when predicting site belonging. According to our results, a major source of the site effect comes from the different scanner models and acquisition protocols, since we achieved the highest accuracy when attempting to classify scanner type (see Suppl. "Harmonization methods").

In addition to scanning differences, demographic and diagnosis distribution were different across the sites. Therefore, we explored if balancing the sample in terms of age and sex distributions would lead to higher classification performance. However, balancing of age/sex distributions across sites did not improve classification performance in Splitting by Site (balanced accuracy 52.6%/50.7% without/with ComBat). Thus, balancing age and sex did not contributed to better performance. As the MDD/HC ratio also varied across sites, an influence of site affiliation to the main MDD vs HC task could exist. Therefore, we additionally explored if the classification performance would drop to random level by equalizing the MDD/HC proportion before splitting the data according to Splitting by Age/Sex. Indeed, we observed a substantial drop of the balanced accuracy from 61% to 53% with 1:1 MDD to HC ratio, confirming our assumption of an incorporation of the site affiliation in the diagnostic classification.

Building on this, ComBat was able to remove the site effect, as all classification models could not differentiate between sites after its application. Subsequently, there were no differences between classification results accross splitting approaches, with around 0.52 in balanced accuracy. Such a low accuracy – close to random chance – is consistent with another large sample study based on two cohorts [17]. In their study, self-reported current depression was speculated as a reason for low accuracy, but this possibility is unlikely explaining our classification results that were based on clinical. Moreover, similar classification levels in our and their study support the notion that a more balanced ratio between classes is not a main aspect behind the low power of discrimination.

Similar to ComBat, other more sophisticated harmonization methods such as ComBat-GAM and CovBat were able to remove site effect, but did not improve the balanced accuracy in Splitting by Site strategy. We cannot exclude the possibility that ComBat-like harmonization tools may overcorrect the data and remove weaker group differences of interest [40]. Hence, encouraging such evaluations in large data sets as well as implementing new methods to be tested [41,42] on both the group and the single subject prediction level could be of great benefit for the imaging community.

*Machine Learning Performance*

In our study, the selection of all classification algorithms was guided by their frequent appearance in prior neuroimaging studies as well as its low computational complexity. According to previous studies [14,17], SVM is the most commonly and successfully used algorithm in previous analyses. We have tested other commonly used ML algorithms, such as logistic regression with LASSO, logistic ridge regression and elastic net logistic [14,43,44]. Given that logistic regression models already have an in-built feature selection procedure, we also included feature selection algorithms such as the two-sample t-test and PCA [45–47], for a fair comparison with SVM. There was no single winner with a significantly higher classification performance across all algorithms, with a balanced accuracy up to 64%, when applied in data split by age/sex, and up to 53%, when split according to subsets of site. A similar trend was observed with AUC. Such a low performance for the Splitting by Site strategy can be a result of simplicity of the models since the highly non-linear interactions of different cortical and subcortical regions are not considered, with an exception of SVM with RBF kernel. In general, specificity was up to 5 % higher than sensitivity, possibly as a result of the imbalanced MDD/HC data sets, even when the impact of both classes was weighted by its ratio during the training.

Considering such a low balanced accuracy, future studies could apply more sophisticated classification methods such as Convolutional Neural Networks [48], which are able to detect nonlinear interactions between all the features as well as to consider spatial information of the given features. As it was demonstrated previously on both real and simulated data [49], regressing out covariates can lead to lower classification performance, therefore one could use an importance weighting instead. Another option would be to include other data modalities such as vertex-wise cortical and subcortical maps [50,51] or even voxel-wise T1 images to capture even more fine-grained changes, which are also present in shapes of subcortical structures[52] or diffusion MRI. A recent resting-state fMRI multi-site study by Qin [53] reported an accuracy of 81.5%. Thus, integration of structural and functional data modalities may result in even higher classification performances.

*Predictive Brain Regions*

Our results do not support the hypothesis that MDD can be discriminated from healthy controls by regional structural features that; classification performance, when site effects were removed, was close to chance level. Nevertheless, during investigation of the most discriminative regions, even after ComBat, we found an overlap with previously reported MDD-related regions. Multiple cortical and subcortical regions were found as the most discriminative between MDD and HC. Most of the cortical regions were identified in previous ENIGMA MDD work [10], which overlaps with our study set of cohorts. Shape differences in left temporal gyrus were reported previously in a younger population with MDD [54]. Left postcentral gyrus and right cuneus surface area were associated with severity of depressive symptoms, while left superior frontal gyrus, bilateral lingual gyrus and left entorhinal cortical thickness were decreased in MDD group [10,55]. In a previous study, MDD subjects exhibited reduced cortical volume compared to HC [56]. Differences in orbitofrontal cortex between MDD and HC were also previously identified [10]. Overall, the effect sizes for case-control differences in these studies were small, which is in line with our current results showing low classification accuracies of these structural brain measures. Additionally, we also found increased thickness of left caudal middle frontal gyrus, right pars triangularis, right superior parietal and right temporal pole in MDD group, which was not previously reported. All subcortical volumes identified as informative for classification became uninformative after ComBat was applied, suggesting that either previously identified alterations in subcortical regions [11] cannot be directly used as MDD predictors at an individual level or ComBat removed differences significant for classification. One possible reason is that subcortical volumes tend to exhibit complex association with the age. Therefore, linear age regression might be an overly simplistic representation of aging trajectories both in ComBat and residualization step. While some of the regions were found also to be predictive in the previous mega-analysis on MDD vs HC study from Stolicyn and colleagues [17], it is dificult to draw a consistent conclusion as they highlight the regions based on the selection frequency by the decision tree model, without reporting the direction of the modulation.

When models were trained and tested only on the subset of features in Splitting by Age/Sex, cortical thicknesses and subcortical volumes yielded higher balanced accuracy compared to cortical surface areas, which is consistent with the previous Enigma MDD meta-analysis, due to an overlap of study cohorts. Yet with the data harmonization step, there was no distinct subgroup of features providing more discriminative information. Together, we observed more changes in weights for cortical thicknesses and subcortical volumes after applying ComBat. One possibility is that differences are more pronounced in thickness than surface area, which is in line with previous findings from univariate approaches [57]. Another possibility is that differences in scanners and acquisition protocols may affect thickness features more strongly than surface areas, in line with previous works [58]. This is a very pertinent topic to be further investigated using multi-cohort mega-analyses on volumetric measures, particularly when the site effect is systematically considered.

Importantly, identified features correspond to the Splitting by Age/Sex strategy as the SVM model was trained on the whole data set with entirely mixed cohorts. While these regions were found to be informative according to the SVM with linear kernel, this model (and every other considered model) failed to differentiate MDD from HC on an individual level, thus one has to be cautious when interpreting these results. Structural alterations in myelination, gray matter, and curvature were found to be associated with MDD-associated genes (Li et al., 2021). Furthermore, a small sample study revealed MDD-related alterations in sulcal depth [59], while white matter topologically-based MDD classification led to up to 76% in accuracy [60]. Thus, the performance could be potentially elevated by integrating morphological shape features with white matter characterestics, such as sulcal depth and curvature, and myelination density as it led to improved performance when classifying sex and autism [61].

*Data stratification*

When the data set was stratified, we found substantial differences in balanced accuracies between the groups only for Splitting by Age/Sex strategy without harmonization step, yet these results were strongly influenced by the site effect. Harmonization step equalizes the accuracies within all pairs of comparisons to a roughly chance probability. Same balanced accuracy was observed when the Splitting by Site strategy was used. This suggests that the demographic and clinical subgroups that we considered do not contain information to predict MDD on an individual level and do not differ in terms of the resultant accuracy, at least according to simplest ML models, despite the group level differences reported prior [10,62]. Large sample meta-analysis of white mater characteristics that investigated similar subgroups also did not reveal significant differences [63], suggesting that the inclusion of these features into ML analysis might not be beneficial for classification improvement. Similarly, a large sample MDD classification study including structural and functional neuroimaging data did not reveal any significant differences between males and females [64]. However, we speculate that other clinically relevant stratifications such as the number of depressive episodes [53,65] and course of disease [53,66] using functional data in further large studies may improve classifications.

## Conclusion

We benchmarked the classification of MDD vs HC using the most commonly used ML algorithms applied to regional surface area features, cortical thickness features and subcortical volumes in the largest multi-site global data set to date. We systematically addressed the questions of general MDD characteristics and power to classify unseen sites by splitting the data in two parallel strategies, which were complemented by ComBat harmonization. A classification accuracy up to 63% was achieved when all cohorts were present in the test set, which decreased down to 52% after ComBat harmonization. Here we have shown that most commonly used ML algorithms may not be able to differentiate MDD from HC on the single subject level using only structural morphometric brain data, even when trained on data from thousands of participants. Furthermore, the performance was not higher in stratified, clinically and demographically more homogeneous groups. Additional work is required to examine if more sophisticated algorithms also known as deep learning can achieve higher predictive power or if other MRI modalities such as task-based or resting state fMRI can provide more discriminative information for successful MDD classification.

## Competing Interest

PMT and NJ received a research grant from Biogen, Inc., for research unrelated to this manuscript. HJG has received travel grants and speakers honoraria from Fresenius Medical Care, Neuraxpharm, Servier and Janssen Cilag as well as research funding from Fresenius Medical Care unrelated to this manuscript. JCS has served as a consultant for Pfizer,Sunovion, Sanofi, Johnson & Johnson, Livanova, and Boehringer Ingelheim. The remaining authors declare no conflict of interest.

## Data availability

Authors are not allowed to share the data of participating sites to third parties inside or outside the ENIGMA MDD consortium. Some sites may provide data upon request.

## Contributions

RGM and VB conceptualized and developed the analysis pipeline, which was approved by ENIGMA MDD working chair LS, co-chair DJV, ENIGMA PI PMT. VB performed all the analyses mentioned in the manuscript and RGM closely supervised them. TEG and EP helped collecting and preparing the data from all participating cohorts. All authors participated in collecting and preprocessing data from their respective sites, reviewed and provided intellectual contribution to the manuscript.

## Acknowledgements

ENIGMA MDD: This work was supported by NIH grants U54 EB020403 (PMT) and R01MH116147 (PMT) and R01 MH117601 (NJ & LS). AMC: supported by ERA-NET PRIOMEDCHILD FP 6 (EU) grant 11.32050.26. AFFDIS: this study was funded by the University Medical Center Goettingen (UMG Startfoerderung) and VB and RGM are supported by German Federal Ministry of Education and Research (Bundesministerium fuer Bildung und Forschung, BMBF: 01 ZX 1507, “PreNeSt - e:Med”). Barcelona-SantPau: MJP is funded by the Ministerio de Ciencia e Innovación of the Spanish Government and by the Instituto de Salud Carlos III through a ‘Miguel Servet’ research contract (CP16–0020); National Research Plan (Plan Estatal de I + D + I 2016–2019); and co-financed by the European Regional Development Fund (ERDF). CARDIFF supported by the Medical Research Council (grant G 1100629) and the National Center for Mental Health (NCMH), funded by Health Research Wales (HS/14/20). CSAN: This work was supported by grants from Johnson & Johnson Innovation (S.E.), the Swedish Medical Research Council (S.E.: 2017–00875, M.H.: 2013–07434, 2019–01138), the ALF Grants, Region Östergötland M.H., J.P.H.), National Institutes of Health (R.D.: R01 CA193522 and R01 NS073939), MD Anderson Cancer Support Grant (R.D.: P30CA016672) Calgary: supported by Canadian Institutes for Health Research, Branch Out Neurological Foundation. FPM is supported by Alberta Children's Hospital Foundation and Canadian Institutes for Health Research. DCHS: supported by the Medical Research Council of South Africa. ETPB: Funding for this work was provided by the Intramural Research Program at the National Institute of Mental Health, National Institutes of Health (IRP-NIMH-NIH; ZIA-MH002857). Episca (Leiden): EPISCA was supported by GGZ Rivierduinen and the LUMC. FIDMAG: This work was supported by the Generalitat de Catalunya (2014 SGR 1573) and Instituto de Salud Carlos III (CPII16/00018) and (PI14/01151 and PI14/01148). Gron: This study was supported by the Gratama Foundation, the Netherlands (2012/35 to NG). Houst: supported in part by NIMH grant R01 085667 and the Dunn Research Foundation. LOND This paper represents independent research (BRCDECC, London) part-funded by the NIHR Maudsley Biomedical Research Centre at South London and Maudsley NHS Foundation Trust and King's College London. The views expressed are those of the authors and not necessarily those of the NHS, the NIHR or the Department of Health and Social Care. MODECT: This study was supported by the Department of Psychiatry of GGZ inGeest and Amsterdam UMC, location VUmc. MPIP: The MPIP Sample comprises patients included in the Recurrent Unipolar Depression (RUD) Case-Control study at the clinic of the Max Planck Institute of Psychiatry, Munich, Germany. We wish to acknowledge Rosa Schirmer, Elke Schreiter, Reinhold Borschke, and Ines Eidner for MR image acquisition and data preparation, and Benno Pütz, and Bertram Müller-Myhsok for distributed computing support and the MARS and RUD Study teams for clinical phenotyping. We thank Dorothee P. Auer for initiation of the RUD study. Melbourne: funded by National Health and Medical Research Council of Australia (NHMRC) Project Grants 1064643 (Principal Investigator BJH) and 1024570 (Principal Investigator CGD). Minnesota the study was funded by the National Institute of Mental Health (K23MH090421; Dr. Cullen) and Biotechnology Research Center (P41 RR008079; Center for Magnetic Resonance Research), the National Alliance for Research on Schizophrenia and Depression, the University of Minnesota Graduate School, and the Minnesota Medical Foundation. This work was carried out in part using computing resources at the University of Minnesota Supercomputing Institute. Moral dilemma: study was supported by the Brain and Behavior Research Foundation and by the National Health and Medical Research Council ID 1125504 to SLW. NESDA: The infrastructure for the NESDA study (www.nesda.nl) is funded through the Geestkracht program of the Netherlands Organisation for Health Research and Development (Zon-Mw, grant number 10–000–1002) and is supported by participating universities (VU University Medical Center, GGZ inGeest, Arkin, Leiden University Medical Center, GGZ Rivierduinen, University Medical Center Groningen) and mental health care organizations, see www.nesda.nl. QTIM: The QTIM data set was supported by the Australian National Health and Medical Research Council (Project Grants No. 496682 and 1009064) and US National Institute of Child Health and Human Development (RO1HD050735). UCSF: This work was supported by the Brain and Behavior Research Foundation (formerly NARSAD) to TTY; the National Institute of Mental Health (R01MH085734 to TTY; K01MH117442 to TCH) and by the American Foundation for Suicide Prevention (PDF-1-064-13) to TCH. SHIP: The Study of Health in Pomerania (SHIP) is part of the Community Medicine Research net (CMR) (http://www.medizin.uni-greifswald.de/icm) of the University Medicine Greifswald, which is supported by the German Federal State of Mecklenburg—West Pomerania. MRI scans in SHIP and SHIP-TREND have been supported by a joint grant from Siemens Healthineers, Erlangen, Germany and the Federal State of Mecklenburg-West Pomerania. This study was further supported by the EU-JPND Funding for BRIDGET (FKZ:01ED1615). SanRaffaele (Milano): Italian Ministry of Health, Grant/Award Number: RF-2011-02349921 and RF-2018-12367489 Italian Ministry of Education, University and Research (Miur). Number: PRIN - 201779W93T. Singapore: The study was supported by grant NHG SIG/15012. KS was supported by National Healthcare Group, Singapore (SIG/15012) for the project.SoCAT: Socat studies supported by Ege University Research Fund (17-TIP-039; 15-TIP-002; 13-TIP-054) and the Scientific and Technological Research Council of Turkey (109S134, 217S228). StanfFAA and StanfT1wAggr: This work was supported by NIH grant R37 MH101495. TIGER: Support for the TIGER study includes the Klingenstein Third Generation Foundation the National Institute of Mental Health K01MH117442 the Stanford Maternal Child Health Research Institute and the Stanford Center for Cognitive and Neurobiological Imaging TCH receives partial support from the Ray and Dagmar Dolby Family Fund.

# Supplementary Materials

Supplementary Table 1: ENIGMA MDD Instrument for diagnosing major depressive disorder and exclusion criteria by site

| Cohort | Diagnosis measurement | Sample characteristics/Inclusion criteria | Exclusion criteria |
|---|---|---|---|
| **AFFDIS** | ICD-10/DSM-IV criteria | MDD subjects currently depressed and in day program or inpatient | All subjects exclusion criteria: current or history of neurological disorder or brain injury, current substance abuse or dependence (not including nicotine), pregnancy, MRI contraindications, inability to give consent. MDD specific: comorbid psychiatric diagnosis. Healthy control specific: current or history of psychiatric diagnosis. |
| **Pharmo (AMC)** | MINI Plus | 48 subjects with lifetime diagnosis of either MDD and/or AD and 14 healthy controls. Patients were startified depending on exposure to SSRIs: early (before age 23) or late (after age 23) exposure to SSRI's, or no exposure at all (UN). 15 subjects were diagnosed with only MDD, 3 with only AD and 22 with both MDD and AD (8 subjects did not receive a diagnosis due to incomplete M.I.N.I. Plus assessment). According to the M.I.N.I. Plus, none of the HC subjects were ever diagnosed with MDD or AD | Less than three week medication-free interval before scanning, current psychotropic medication use, a history of chronic or neurological disorder, family history of sudden heart failure or epileptic attacks, pregnancy (tested via urine sampling prior to the assessment), breast feeding, alcohol dependence and contra-indications for an MRI scan (e.g., ferromagnetic fragments). Participants agreed to abstain from smoking, caffeine and alcohol use for 24 hours prior to the assessments. |
| **Barcelona** | DSM-IV-TR acc. to CIDI-interview and HAMD | Outpatients with MDD diagnosis (DSM-IV-TR), with a first episode, recurrent MDD or chronic MDD (TRD) age 18-65 | The exclusion criteria for healthy participants were: lifetime psychiatric diagnoses, first-degree relatives with psychiatric diagnoses and clinically significant physical or neurological illnesses. Axis I comorbidity according to DSM-IV-TR criteria was an exclusion criteria for all participants. |
| **Cardiff** | Hamilton Depression Rating Scale (HDRS-17) | N= 40, MDD patients with a current moderate to severe depressive episode despite minimum three months of stable antidepressant treatment | Psychotic symptoms, current substance dependence, eating disorders, claustrophobia and other MRI contraindications, and ongoing non-pharmacological treatment. |
| **CSAN (Adf)** | MINI | Current MDD: Meets MINI criteria for depression; comorbid anxiety disorders are allowed; mood-congruent psychotic symptoms allowed. | Current MDD: a current DSM-5 diagnosis of substance use disorder, except nicotine; a psychotic disorder, except depression with mood-congruent psychotic features; new antidepressant medication during the month before study participation (two months for fluoxetine); change of the dose of psychotropic medications over the last month (antidepressant and antipsychotic medication) or the last two months (mood stabilizers and anticonvulsants). |
| **Calgary** | KSADS | First episode MDD and healthy controls (Dalhousie sample). Recurrent MDD and healthy controls, recruited via referral from clinicians in Calgary, Alberta and through advertisements in local clinics and at the University of Calgary (Calgary sample). | Dalhousie Sample: A history of neurological illness, medical illness, claustrophobia, >21 year of age, or the presence of a ferrous implant or pacemaker. University of Calgary: Left handed; history of seizures, epilepsy or other neurological or psychiatric diagnoses (specifically bipolar disorder, psychosis, pervasive developmental disorder, eating disorders, PTSD); pregnancy |
| **DCHS** | MINI | Women over the age of 18 years, who were between 20 and 28 weeks pregnant, who presented at either of the two recruitment clinics, and who had no intention of moving out of the area within the following year, and were able to give written consent | 1) loss of consciousness longer than 30 minutes, 2) inability to speak English, 3) current/lifetime alcohol and/or substance dependence or abuse, 4) psychopathology other than PTSD and/or MDD, 5) traumatic brain injury, 6) standard MRI exclusion criteria |
| **ETPB** | HAMD,BDI, SHAPS,MADRS | Treatment resistant depression, at least one failed trial MADRS >20 | Current or past diagnosis of Schizophrenia or any other psychotic disorder as defined in the DSM-IV. Subjects with a history of DSM-IV drug or alcohol dependency or abuse (except |
| | | | for nicotine or caffeine) within the preceding 3 months. Female subjects who are either pregnant or nursing. Serious, unstable illnesses including hepatic, renal, gastroenterologic, respiratory, cardiovascular (including ischemic heart disease), endocrinologic, neurologic, immunologic, or hematologic disease. Subjects with uncorrected hypothyroidism or hyperthyroidism. Subjects with one or more seizures without a clear and resolved etiology. Treatment with a reversible MAOI within 4 weeks prior to study phase I. Treatment with fluoxetine within 5 weeks prior to study phase I. Treatment with any other concomitant medication not allowed (Appendix A for Substudy 2; Appendix G for Substudy 4) 14 days prior to study phase I. No structured psychotherapy will be permitted during the study. Current NIMH employee/staff or their immediate family member. Additional Exclusion Criteria for substudy 2 (patients with MDD) Previous treatment with ketamine or hypersensitivity to amantadine. Additional Exclusion Criteria for Substudy 4 (patients with MDD or BD). Subjects who currently are using drugs (except for caffeine or nicotine), must not have used illicit substances in the 2 weeks prior to screen and must have a negative alcohol and drug urine test (except for prescribed benzodiazepines) urine test at screening. Presence of any medical illness likely to alter brain morphology and/or physiology (e.g., hypertension, diabetes) even if controlled by medications. Clinically significant abnormal laboratory tests. Presence of metallic (ferromagnetic) implants (e.g, heart pacemaker, aneurysm clip). Subjects who, in the investigator s judgment, pose a current serious suicidal or homicidal risk, or who have a MADRS item 10 score of >4. |
| **EPISCA (Leiden )** | ADIS | Inclusion criteria for the patient group were: having clinical depression as assessed by categorical and dimensional measures of DSM-IV depressive and anxiety disorders, no current and prior use of antidepressants, and being referred for CBT at an outpatient care unit. Inclusion criteria for the control group were: no current or past DSM-IV classifications, no clinical scores on validated mood and behavioral questionnaires, no history of traumatic experiences, and no current psychotherapeutic and/or psychopharmacological intervention of any kind. | Primary DSM-IV clinical diagnosis of ADHD, ODD, CD, pervasive developmental disorders, post-traumatic stress disorder, Tourette's syndrome, obsessive–compulsive disorder, bipolar disorder, and psychotic disorders; current substance abuse; history of neurological disorders or severe head injury; age < 12 or > 21 years; pregnancy; left-handedness; IQ score < 80 as measured by the Wechsler Intelligence Scale for Children (WISC) (Wechsler, 1991) or Adults (Wechsler, 1997); and general MRI contra- indications. |
| **FIDMA G** | DSM-IV-TR criteria | MDD patients within a current depressive episode (HDRS >= 17, only 1 patient was in remission), right-handed, age 18-65 | Patients were excluded (i) if they were left-handed; (ii) if they were younger than 18 or older than 65 years; (iii) if they had a history of brain trauma or neurological disease; (iv) if they had shown alcohol/ substance abuse within 12 months prior to participation; and (v) if they had undergone electroconvulsive therapy in the previous 12 months. |
| **Groningen sample (DIP)** | MINI-SCAN | Outpatients with MDD diagnosis. Inclusion MDD: Outpatients treated in mental health care for depression, BDI-II>13 at screening, adults. | Exclusion MDD: Comorbid axis-I disorders other than anxiety disorders or past substance abuse, other psychotropic medication than stable use of SSRI/SNRI/TCA, established cardiovascular disease, active and concrete suicidal plans, inadequate language proficiency, cognitive impairments or neurological disease that interferes with task performance. Exclusion CTL: Same as MDD, lifetime history of MDD, BDI>8. |
| **Houston** | SCID interview | Outpatients | MDD subjects: age below 18; lifetime or current diagnosis of psychotic disorder, or bipolar I or II disorder; substance abuse/dependence in 6 months prior to study inclusion; current major medical problems. Control subjects: age below 18; current major medical problems; current psychiatric or neurologic disorder; history of psychiatric disorders in a first-degree relative; current major medical problems. Both groups: MRI contra-indications |
| **TiPs (Jena, Germany)** | SCID interview | Psychiatric inpatients and tinnitus patients with MDD or a disorder of the depressive spectrum (also adjustment disorders as pointed out in the data table); psychiatrically healthy controls were derived from community and tinnitus patients | MDD subjects: presence of axis-I disorders other than MDD or adjustment disorders. Control subjects: no Axis-I diagnosis, no medication use. Exclusion criteria for all subjects included history of neurological disease (e.g. tumour, head trauma, epilepsy) or untreated internal medical condtitions, intellectual and/or developmental disability. Only German native speakers were allowed to participate. |
| **BRCDECC London** | SCAN interview | Community based or outpatients, none were inpatients. MDD subjects: Less than two depressive episodes of at least moderate severity. Did not meet DSM-IV diagnostic criteria for recurrent major depressive disorder.<br>Control group participants were clinically interviewed to ensure they had never experienced depressive symptoms. Exclusion criteria for all participants were for contraindications to MRI; other exclusion criteria were a diagnosis of neurological disorder, head injury leading to loss of consciousness or conditions known to affect brain structure or function (including alcohol or substance misuse), ascertained during clinical interview. Potential participants were also excluded if they or a first-degree relative had ever fulfilled criteria for mania, hypomania, schizophrenia or mood-incongruent psychosis. | Contraindications to MRI, diagnosis of neurological disorder, head injury leading to loss of consciousness or conditions known to affect brain structure or function (including alcohol or substance misuse), if they or a first-degree relative had ever fulfilled criteria for mania, hypomania, schizophrenia or mood-incongruent psychosis. |
| **MODECT** | MINI | Older adults, aged above 55, with severe depression admitted to be treated with ECT | Exclusion criteria were another major DSM-IV-TR diagnosis, such as schizophrenia, bipolar or schizoaffective disorder and a history of major neurological illness (including Parkinson’s disease, stroke and dementia). |
| **MPIP** | M-CIDI/SCAN interview | M. A. R. S. sample: both first and recurrent episodes; RUD sample: only recurrent episodes with some patients scanned in remission | 1. Munich Antidepressant Response Signature (MARS) study MDD subjects (clinical consensus diagnosis or M-CIDI (since 2008)): depressive syndromes secondary to any medical or neurological condition (e. g., intoxication, drug abuse, stroke), the presence of manic, hypomanic or mixed affective symptoms, lifetime diagnosis of alcohol dependence, illicit drug abuse or the presence of severe medical conditions (e.g., ischemic heart disease). Patients with bipolar depression were excluded for the current MR study. Control subjects: age > 65, MMSE<27, presence of severe somatic diseases or lifetime history of the following axis I disorders as assessed by the M-CIDI interview: alcohol dependence, drug abuse or dependence, possible psychotic disorder, mood disorder, anxiety disorder including OCD and PTSD, somatoform disorder, dissociative disorder NOS, and eating disorder 2. Recurrent unipolar depression (RUD) study: MDD subjects (SCAN interview): presence of manic episodes, mood incongruent psychotic symptoms, the presence of a lifetime diagnosis of intravenous drug abuse and depressive symptoms only secondary to alcohol or substance abuse or to medical illness or medication.Control subjects: presence of |
| | | | severe somatic diseases or life-time history of anxiety and affective disorders according to the Composite International Diagnostic-Screener (CIDI-S). All subjects: gross incidental MR findings such as territorial infarction, tumor, hydrocephalus, malformations and anatomical deviations (e.g. enlarged ventricles) that prevent appropriate image processing were additional exclusion criteria. 3. MR images of 9 additional controls acquired at the LMU, Munich, meeting equivalent criteria as the RUD control sample were included. |
| **Melbourne** | SCID interview | Youth depression sample: 15-25 years of age. Recruited as part of 2 large RCTs (incl. YoDA-C - Davey et al., 2014; Trials) and scanned prior to treatment randomisation. 60 patients unmedicated (YoDA-C). | MDD subjects: lifetime or current SCID-I diagnosis of psychotic disorder, or bipolar I or II disorder. Control subjects: any SCID-I diagnosis or medication use. Both groups: Acute or unstable medical disorder; general MRI contraindications |
| **Minnesota** | Schedule for Affective Disorders and Schizophrenia for School-Age Children–Present and Lifetime Version and the Children's Depression Rating Scale–Revised (CDRS-R). | Adolescents with MDD and HC aged 12 to 19 years were recruited to participate through community postings and referrals from local mental health services. Adolescents with MDD were eligible if they had a primary diagnosis of MDD and had not received any psychotropic medication treatment for the past 2 months. Healthy adolescents were eligible if they had no current or past psychiatric diagnoses and were frequency matched to the MDD group on age and sex | Exclusion criteria for both groups included the presence of a neurologic or other chronic medical condition, mental retardation, pervasive developmental disorder, substance use disorder, bipolar disorder, or schizophrenia |
| **Moral Dilemma** | SCID interview | Youth depression sample: 15-25 years of age; recruited from outpatient service. Controls recruited from general community. | MDD subjects: lifetime or current SCID-I diagnosis of psychotic disorder, or bipolar I or II disorder; current antidepressant medication use. Control subjects: any SCID-I diagnosis or medication use. Both groups: Acute or unstable medical disorder; general MRI contraindications |
| **NESDA** | CIDI interview | DSM-4 based diagnosis of MDD (6 month recency), using CIDI interview. 93 (60%) MDD patients have a comorbid ANX diagnosis. Age range 18-65 | N/A |
| **QTIM** | CIDI interview | Retrospective questionnaire about depression episodes combined with an MRI study. The best described MDD episode is defined as the worst one (according to individuals). We have up to 5 supplementary episodes (briefly) described. Sample composed of twins and relatives. Population-based sample | MDD subjects: presence of axis-I disorders other than MDD and anxiety disorders Control subjects: antidepressant use, psychiatric disorders All subjects: relatedness between subjects, left handedness, history of neurological or other severe medical illness, head injury or current or past diagnosis of substance abuse, use of cognition affecting medication and general MRI contraindications |
| **San Francisco UCSF** | KSADS (semi-structured interview based on DSM) for MDD, DISC/DPS for HCL | Outpatient/community-based sample with DSM diagnosis, mostly antidepressant-naive and approximately 60% of MDD have comorbid anxiety disorders | Exclusion criteria for all participants included: 1) use of pharmacotherapeutics for treating psychiatric conditions within the past 6 months, 2) misuse of drugs within two months prior to MRI scanning; 3) two or more alcoholic drinks per week within the previous month (as assessed by the Customary Drinking and Drug Use Record; CDDR) (Brown et al, 1998); 4) a full scale IQ score of less than 75 (as assessed by the Wechsler Abbreviated Scale of Intelligence; WASI) (Wechsler, 1999); 5) contraindications for MRI including ferromagnetic implants and claustrophobia; 6) pregnancy or the possibility of pregnancy; 7) left-handedness; 8) prepubertal status (as assessed as Tanner stages of 1 or 2) (Tanner, 1962); 9) inability to understand and comply with procedures; 10) neurological disorder (including meningitis, migraine, or HIV); 11) head trauma; 12) learning disability; 13) serious health problems; and 14) complicated or premature birth (i.e., birth before 33 weeks of gestation). The MDD group was subject to |
| | | | the additional exclusion criterion of a primary psychiatric diagnosis other than MDD. The HCL group was subject to the additional exclusion criteria of: 1) history of mood or psychotic disorders in a first- or second-degree relative (as assessed by the Family Interview for Genetics; FIGS) (Maxwell, 1992); and 2) current or lifetime DSM-IV-TR Axis I psychiatric disorder. |
| **SHIP** | M-CIDI interview | Population based longitudinal cohort study | MDD subjects: presence of axis-I disorders other than MDD, anxiety disorders, conversion, somatization and eating disorder. Control subjects: no lifetime diagnosis of depression, no antidepressiva, and severity index=0 All subjects: We removed subjects with medical conditions (e.g. a history of cerebral tumor, stroke, Parkinson's diseases, multiple sclerosis, epilepsy, hydrocephalus, enlarged ventricles, pathological lesions) or due to technical reasons (e.g. severe movement artifacts or inhomogeneity of the magnetic field). |
| **SHIP-TREND** | M-CIDI interview | Population based longitudinal cohort study | MDD subjects: no special exclusion criteria Control subjects: no lifetime diagnosis of depression, no antidepressiva, and severity index=0 All subjects: We removed subjects with due to medical conditions (e.g. a history of cerebral tumor, stroke, Parkinson's diseases, multiple sclerosis, epilepsy, hydrocephalus, enlarged ventricles, pathological lesions) or due to technical reasons (e.g. severe movement artifacts or inhomogeneity of the magnetic field). |
| **San Raffaele Milano OSR** | SCID interview | adult MDD depressed inpatients | Other diagnoses on Axis I; pregnancy, history of epilepsy, major medical and neurological disorders; absence of a history of drug or alcohol dependency or abuse within the last six months. inflammation-related symptoms, including fever and infectious or inflammatory disease; uncontrolled systemic disease; uncontrolled metabolic disease or other significant uncontrolled somatic disorder known to affect mood; somatic medications known to affect mood or the immune system, such as corticosteroids, non-steroid anti-inflammatory drugs and statins. |
| **Singapore** | SCID interview | Inclusion: 1) DSM IV dx of MDD (Patients) 2) Age: 21-65 3) English speaking 4) Provision of informed written consent | Exclusion criteria 1) History of significant head injury 2)Neurological diseases such as epilepsy, cerebrovascular accident 3) Impaired thyroid function 4) Steroid use 5) DSM IV alcohol or substance use or dependence 6) Contraindications to MRI (e.g. pacemaker, orbital foreign body, recent surgery/procedure with metallic devices/implants deployed) using standard MRI Request Form from NNI 7)Pregnant women 8) Claustrophobia |
| **SoCAT** | SCID interview | Inclusion criteria: DSM IV dx for mdd patients Age: 18-65 right-handed currently depressed or remitted; Control subjects: any history of psychiatric disorder | Exclusion criteria 1) History of significant head injury 2)Neurological diseases such as epilepsy, cerebrovascular accident 3)Other diagnoses on Axis I disorders4) |
| **Stanford FAA** | SCID interview | Community-based DSM-diagnosed sample | MDD subjects: presence of axis-I disorders other than MDD, anxiety and eating disorders . Control subjects: control individuals did not meet diagnostic criteria for any current psychiatric. Both groups: alcohol / substance abuse or dependence within six months prior to MRI scanning, history of head trauma with loss of consciousness > 5 min, aneurysm, or any neurological or metabolic disorders that require ongoing medication or that may affect the central nervous system (including thyroid disease, diabetes, epilepsy or other seizures, or multiple sclerosis), MRI contraindications, or bad MRI data (e.g., extreme movement). |
| **Stanford T1w** | SCID interview | Community-based DSM-diagnosed sample | MDD subjects: presence of axis-I disorders other than MDD, anxiety and eating disorders . |
| **Aggregate** | | | Control subjects: control individuals did not meet diagnostic criteria for any current psychiatric. Both groups: alcohol / substance abuse or dependence within six months prior to MRI scanning, history of head trauma with loss of consciousness > 5 min, aneurysm, or any neurological or metabolic disorders that require ongoing medication or that may affect the central nervous system (including thyroid disease, diabetes, epilepsy or other seizures, or multiple sclerosis), MRI contraindications, or bad MRI data (e.g., extreme movement). |
| **TIGER** | KSADS | Community-based DSM-diagnosed sample | All subjects: Exclusion criteria were premenarchal status (for females), history of concussion within the past 6 weeks or history of any lifetime concussion with loss of consciousness, contraindications to MRI scanning (e.g. braces, metal implants, or claustrophobia), serious neurological or intellectual disorders that could interfere with the participant's ability to complete study components. MDD subjects: meeting lifetime or current DSM-IV criteria for any Bipolar Disorder, Psychosis, or Alcohol Dependence, or DSM-5 criteria for Moderate Substance Use Disorder with substance-specific threshold for withdrawal. CTL subjects: any current or past DSM-IV Axis I Disorder and first-degree relative with confirmed or suspected history of depression, mania, psychosis, or substance dependence. |

Supplementary Table 2: ENIGMA MDD Image acquisition and processing by cohort

| Cohort | Scanner type | Sequence T1 | FreeSurfer version | Slice orientation | Operating system |
|---|---|---|---|---|---|
| **AFFDIS** | 3T Siemens Magnetom TrioTim | 3D T1 (176 slices; TR = 2250 ms; TE = 3.26 ms; FOV 256; voxel size 1X1X1mm) | 5,3 | Sagittal | Linux CentOS |
| **Pharmo (AMC)** | 3T Philips | T1 sequence details: 3D-TFE sequence TR= 9.7 ms, TE=4.6ms, matrix 192x192, voxel size = 0.875 x 0.875 x 1.2 mm; 120 slices. Axial plane. Philips 3T Ingenia 16 channel coil | 5,3 | Transverse | freesurfer-Linux-centos6_x86_64-stable-pub-v5.3.0 |
| **Barcelona** | 3T Philips Achieva | 3D MPRAGE images (Whole-brain T1-weighted); TR=6.7ms, TE=3.2ms; 170 slices, voxel size 0.89X0.89X1.2 mm. Image dimensions 288X288X170; field of view: 256X256X204; slice thickness: 1.2 mm; with a sagittal slice orientation, T1 contrast enhancement, flip angle: 8º, grey matter as a reference tissue, ACQ matrix MXP = 256X240 and turbo-field echo shots (TFE) = 218. | 6 | Sagittal | Scientific Linux 5 |
| **Cardiff** | A 3 Tesla whole body MRI system (General Electric, Milwaukee, USA) with an 8-channel head coil was used at the Cardiff University | High-resolution anatomical scan (Fast Spoiled Gradient-Recalled-Echo [FSPGR] sequence): 178 slices, TE=3 ms, TR=7.9 ms, voxel size=1.0×1.0×1.0 mm3, FA=15°, FOV=256×256 | 5,3 | | freesurfer-Linux-centos6_x86_64-stable-pub-v5.3.0 |
| | Brain Research Imaging Centre (CUBRIC). | | | | |
| **CSAN (Adf)** | 3T Siemens MAGNETOM PRISMA | Whole-head t1-weighted MPRAGE (TR = 2300 ms, TE = 2.34 ms, FOV 250 × 250 mm, voxel size = 0.9 × 0.868 × 0.868 mm, flip angle = 8°) | 7.2 | Sagittal | Ubuntu |
| **Calgary** | 1.5T Siemens Magnetom Vision. 3T GE Discovery MR750 | 1.5T: A sagittal scout series was acquired to test image quality. 3D fast low angle shot (FLASH) sequence was used to acquire data from 124 1.5 mm-thick contiguous coronal slices through the entire brain (echo time = 5ms, repetition time = 25ms, acquisition matrix = 256 x 256 pixels, field of view = 24 cm and flip angle = 40°). 3T: Anatomical imaging acquisition parameters: axial acquisition, repetition time (TR), 2200 milliseconds (ms); echo time (TE), 3.04 ms; TI, 766, 780; flip angle, 13 degrees; 208 partitions; 256 × 256 matrix; and field of view, 256. | 5,3 | Dalhousie sample, coronal; Calgary sample, axial | MacOs Sierra |
| **DCHS** | 3T Siemens Skyra | 3D multi-echo MPRAGE, voxel size 1 mm x 1mm x 1.5mm, TR = 2530 ms, TE = 1.69 x 3.55 x 5.41 x 7.27ms, FOV: 256x256mm, flip angle = 7° | 5.3 | Sagittal | Linux-centos6_x86_64 |
| **ETPB** | 3T, GE HDx | Fast spoiled gradient recalled echo (FSPGR). Slice Thickness: 1. Repetition Time: 8.836. Echo Time: 3.496. Inversion Time: 450. Magnetic Field Strength: 3. Spacing Between Slices: 1. Echo Train Length: 1. Percent Sampling: 100. Percent Phase Field of View: 100. Pixel Bandwidth: 195.312. Reconstruction Diameter: 256. Acquisition Matrix: 256x256. In-plane Phase Encoding Direction: ROW. Flip Angle: 13 | 5,3 | Sagittal | Linux |
| **EPISCA (Leiden)** | 3T Philips Achieva | a sagittal 3-dimensional gradient-echo T1-weighted image was acquired (repetition time = 9.8 ms; echo time = 4.6 ms; flip angle = 8°; 140 sagittal slices; no slice gap; field of view =256 × 256 mm; 1.17 × 1.17 × 1.2 mm voxels; duration = 4:56 min) | 5,3 | Sagittal | Ubuntu 14.04.5 LTS (Linux 3.13.0-153-generic x86_64) |
| **FIDMAG** | 1.5T, GE Signa | 3D T1: matrix size = 512 × 512, 180 contiguous axial slices, voxel resolution = 0.47 × 0.47 × 1mm, no slice gap, TE = 3.93ms, TR = 2000ms and inversion time (TI) = 710ms, flip angle = 15 degrees | 6 | Axial | Linux-centos6_x86_64 |
| **Groningen sample (DIP)** | 3T Philips | 3D T1-weighted scan (170 slices; TR = 9ms; TE = 3.6ms; 256x231 matrix of 1×1×1 mm voxels) | 5,3 | Sagittal | SUSE Linux X86_64 |
| **Houston** | subjects in 20000s: 1.5 T Philips Medical Systems Gyroscan Intera; subjects in 30000s: 3T Siemens Allegra | Subjects in the 20000s: Fast field echo sequence- repetition time (TR) = 24 ms, echo time (TE) = 4.99 ms, flip angle = 40°, slice thickness = 1 mm, matrix size = 256 × 256 and 150 slices. Subjects in 30000s: MPRAGE-repetition time (TR) = 1750 ms, echo time (TE) = 4.39 ms, flip angle = 8°, slice thickness = 1 mm, matrix size = 208 × 256 and 160 slices. | 5,3 | Subjects in 20000s: Sagittal; Subjects in 30000s: Transverse | Fedora 19 |
| **TiPs (Jena, Germany)** | 3T Siemens MAGNETOM Prisma_fit | MPRAGE sequence: TR 2300 ms, TE 3.03 ms, α 9°, 192 contiguous sagittal slices, in-plane field of view 256 mm, voxel resolution 1x1x1 mm; acquisition time 5:21 min | 5,3 | Sagittal | Linux |
| **BRCDECC London** | 1.5T GE Signa HDx | ADNI-1 MPRAGE pulse sequence (details at | 5,3 | Sagittal | Linux-centos4_x86_64 |
| | | http://adni.loni.ucla.edu/research/protocols/mri-protocols/) | | | |
| **MODECT** | 3T (General Electric Signa HDxt, Milwaukee, WI, USA) | T1-weigthed data set was acquired (flip angle=12°, repetition time=7.84 milliseconds, echo time=3.02 milliseconds; matrix 256x256, voxel size 0.94x0.94x1 mm; 180 slices). | 5,3 | Coronal | Linux |
| **MPIP** | 1.5T GE and Siemens (the latter: only few cases) | #1: T1-weighted SPGR sagittal 3D volume. TR=1030 msec; TE=3.4 msec; 124 slices; matrix=256x256; FOV=23.0x23.0 cm2; voxel size=0.8975 mm x0.8975 mm x 1.2-1.4 mm; flip angle=90°; birdcage resonator. #2: same scanner as #1, platform update Signa Excite, sagittal T1-weighted (spin echo sequence, TR=9.7 msec, TE=2.1 msec; FOV=25.0x25.0 cm2, voxel size=0.875 mm x0.875 mm x1.2 mm, 124- 132 slices, flip angle=90°. #3: Siemens 1.5 Tesla, Vario, 3D MPRAGE, TR=11.6 msec; TE=4.9 msec; FOV 23x23 cm2; matrix 512x512; 126 axial slices; voxel site 0.45 mm x 0.45 mm x 1.5 mm. (only N=2 subjects) | 5,3 | 1.5 GE: sagittal. 1.5 Siemens: axial | Linux 2.6.37.1-1.2- desktop x86_64 |
| **Melbourne** | 3T GE Signa Excite | 3D BRAVO sequence 140; TR=7900 ms; TE=3000 ms; flip angle=13º; FOV=256 mm; matrix=256 x 256 | 5,3 | Axial | Linux Debian x86 64 |
| **Minnesota** | 3.0 Tesla Tim Trio scanner; Siemens Corp | A 5-minute structural scan was acquired using a T1-weighted, high-resolution, magnetization-prepared gradient-echo sequence: repetition time, 2530 milliseconds; echo time, 3.65 milliseconds; inversion time, 1100 milliseconds; flip angle, 7°; field of view, 256 × 176 mm; voxel size, 1-mm isotropic; 224 slices; and generalized, autocalibrating, partially parallel acquisition acceleration factor, 2. | 5,3 | Coronal | Linux |
| **Moral Dilemma** | 3T GE Signa Excite | 3D BRAVO sequence: 140 contiguous slices; repetition time, 7900 ms; echo time, 3000 ms; flip angle, 13°; in a 25.6-cm field of view, with a 256 × 256 pixel matrix and a slice thickness of 1 mm (1 mm gap). | 5,3 | Axial | Linux Debian x86 64 |
| **NESDA** | 3T Phillips Achieva/Intera | 3D gradient-echo T1-weighted sequence. TR=9 msec; TE=3.5 msec; flip angle 8º, FOV = 256 mm; matrix: 25x62x56; in plane voxel size = 1 mm × 1 mm x 1 mm; 170 slices. | 5 | Sagittal | SHARK HPC, Linux environment |
| **QTIM** | Bruker 4T Wholebody MRI | 3D T1 weighted sequence. TR=1500 msec; TE=3.35 msec; flip angle=8°, 256 or 240 (coronal or sagittal) slices, FOV=240 mm, matrix 256x256x256 (or 256x256x240) | 5,1 | Coronal, then sagittal following software upgrade. | Linux-centos4_x86_64-stable-pub-v5.1.0 |
| **San Francisco UCSF** | 3T GE Discovery MR750 | SPGR T1-weighted: TR=8.1 ms; TE=3.17 ms; TI=450 ms; flip angle=12°; 256x256 matrix; FOV=250x250 mm; 168 sagittal slices; slice thickness=1 mm; in-plane resolution=0.98x 0.98 mm | 5,3 | Sagittal | Linux-centos6_x86_64-stable-pub-v5.3.0. |
| **SHIP** | 1.5T Siemens Avanto | 3D T1-weighted (MP-RAGE/ axial plane); TR=1900 msec; TE=3.4 msec; Flip angle=15°; voxel size 1 mm x 1 mm x 1 mm | 5.3 (cortical), 5.1 (subcortical) | Axial | Centos6_x86_64 |
| **SHIP-TREND** | 1.5T Siemens Avanto | 3D T1-weighted (MP-RAGE/ axial plane); TR=1900 msec; TE=3.4 msec; Flip angle=15°; voxel size 1 mm x 1 mm x 1 mm | 5.3 (cortical), 5.1 (subcortical) | Axial | Centos6_x86_64 |
| **San Raffaele Milano OSR** | 3T Philips Ingenia and 3T Philips Intera scanner | 3D-MPRAGE sequence: TR 2500 ms, TE 4.6 ms, field of view FOV = 230 mm, matrix = 256 × 256, in-plane resolution 0.9 × 0.9 mm, yielding 220 transversal slices with a thickness of 0.8 mm. | 5,3 | axial | Linux Ubuntu 16.04 |
| **Singapore** | Achieva 3T, Philips Medical Systems, Netherlands | Whole brain high resolution 3D MP-RAGE (magnetisation-prepared rapid acquisition with a gradient echo) volumetric scans (TR/TE/TI/flip angle 8.4/3.8/3000/8; matrix 256x204; FOV 240mm2) with axial orientation (reformatted to coronal) | 5,3 | Axial | Linux_Ubuntu12.04_6 4 |
| **SoCAT** | 3.0 T, Siemens Verio,Numaris/4,Syngo MR B17,Erlangen ,Germany | 3D T1 weighted MP-Rage/axial plane; TR=1900 msec; TE=3.4 msec; Flip angle=15°; Voxel size 1 mm x 1 mm x 1 mm | 5,3 | Axial | Ubuntu 18.04 LTS |
| **Stanford FAA** | 3.0T GE Discovery MR750 | Whole-brain T1-weighted images were collected using a spoiled gradient echo (SPGR) pulse sequence (186 sagittal slices; resolution = 0.9 mm isotropic; flip angle = 12°; repetition time [TR] = 6,240 ms; echo time [TE] = 2.34 ms) | 5,3 | Sagittal | Linux-centos6_x86_64 |
| **Stanford T1w Aggregate** | 1.5T GE Signa Excite | Whole-brain T1-weighted images were collected using a spoiled gradient echo (SPGR) pulse sequence (116 sagittal slices; through-plane resolution = 1.5 mm; in-plane resolution = 0.86 x 0.86 mm; flip angle = 15 degrees; repetition time [TR] = 8.3-10.1 ms; echo time [TE] = 1.7-3.0; inversion time [TI] = 300 ms; matrix = 256 x 192). | 5,3 | Sagittal | Centos6_x86_64, Linux-based HPC |
| **TIGER** | 3T GE MR750 | TR/TE/TI=8.2/3.2/600 ms; flip angle=12°; 156 axial slices; FOV=25.6 cm; matrix=256 mm x 256 mm, isotropic voxel=1 mm, total scan time: 3:40 | 6 | Axial | Linux |

Supplementary Table 3: List of hyperparameters of trained algorithms. Optimal hyperparameters were found by the grid search. We followed a heuristic approach outlined in [1] to determine a range of values for C and $\gamma$. To access different power of regularization, we followed the same range of values for $\lambda$. The range of values of random forests hyperparameters were optimized according to [2].

| Classification algorithm | Feature Selection | Hyperparameters | Nested CV |
|---|---|---|---|
| **SVM Linear** | None | C = $[10^{-4}, 10^{-3}, \dots, 10^{4}]$ | 10 fold |
| **SVM Linear** | PCA | C = $[10^{-4}, 10^{-3}, \dots, 10^{4}]$<br><br>% components = [10%,20%, ... ,100%] | 10 fold |
| **SVM Linear** | Ttest (p-value<0.05) | C = $[10^{-4}, 10^{-3}, \dots, 10^{4}]$ | 10 fold |
| **SVM rbf** | None | C = $[10^{-4}, 10^{-3}, \ldots, 10^{4}]$<br><br>$\gamma = [10^{-4}, 10^{-3}, \ldots, 10]$ | 10 fold |
| **LASSO** | None | $\lambda = [10^{-4}, 10^{-3}, \ldots, 10^{4}]$ | 10 fold |
| **Ridge** | None | $\lambda = [10^{-4}, 10^{-3}, \ldots, 10^{4}]$ | 10 fold |
| **Elastic Net** | None | $\lambda = [10^{-4}, 10^{-3}, \ldots, 10^{4}]$<br><br>$\alpha = [0.1, 0.2, \ldots, 1]$ | 10 fold |
| **Random Forests** | None | max depth = [40,50,...,100]<br><br>number of trees = [400,600,..., 1600]<br><br>min samples split = [5,10,.., 25]<br><br>number of random features = [sqrt, None ] | 10 fold |

Supplementary Table 4: Clinical sample characteristics. Number of subjects with major depressive disorder (MDD) in each category.

| | Number of episodes | | Antidepressant (AD) use | | Age of Onset | |
|---|---|---|---|---|---|---|
| **Cohorts** | First episode | Recurrent episodes | no AD | with AD | Adolescent | Adults |
| **AFFDIS** | 3 | 30 | 1 | 32 | 61 | 18 |
| **AMC** | 20 | 22 | 50 | 0 | 24 | 24 |
| **Barc** | 22 | 40 | 4 | 58 | 13 | 49 |
| **CARDIFF** | 0 | 35 | 0 | 40 | 11 | 22 |
| **CSAN** | 14 | 46 | 31 | 29 | 0 | 0 |
| **Calgary** | 18 | 49 | 49 | 19 | 47 | 1 |
| **DCHS** | 0 | 0 | 0 | 0 | 0 | 0 |
| **ETPB** | 0 | 34 | 34 | 0 | 28 | 5 |
| **Episca** | 19 | 0 | 18 | 1 | 0 | 0 |
| **FIDMAG** | 11 | 22 | 4 | 30 | 6 | 27 |
| **Gron** | 6 | 12 | 10 | 10 | 8 | 11 |
| **Houst** | 41 | 50 | 101 | 1 | 50 | 25 |
| **Jena** | 7 | 23 | 12 | 19 | 0 | 0 |
| **LOND** | 0 | 69 | 19 | 50 | 36 | 18 |
| **MODECT** | 0 | 33 | 32 | 10 | 0 | 0 |
| **MPIP** | 91 | 246 | 53 | 284 | 71 | 266 |
| **Melb** | 48 | 89 | 121 | 22 | 126 | 4 |
| **Minnesota** | 16 | 22 | 52 | 16 | 66 | 0 |
| **Moraldilemma** | 8 | 16 | 24 | 0 | 0 | 0 |
| **NESDA** | 67 | 87 | 98 | 56 | 78 | 76 |
| **QTIM** | 0 | 0 | 73 | 29 | 87 | 15 |
| **SF** | 32 | 34 | 75 | 0 | 60 | 0 |
| **SHIP_S2** | 77 | 59 | 113 | 23 | 461 | 118 |
| **SHIP_T0** | 113 | 197 | 257 | 53 | 974 | 255 |
| **SanRaffaele** | 1 | 44 | 2 | 42 | 7 | 37 |
| **Singapore** | 8 | 14 | 4 | 18 | 2 | 20 |
| **Socat_dep** | 19 | 60 | 41 | 38 | 121 | 28 |
| **StanfFAA** | 0 | 14 | 11 | 3 | 11 | 3 |
| **StanfT1wAggr** | 6 | 48 | 27 | 20 | 36 | 18 |
| **TIGER** | 29 | 20 | 29 | 20 | 49 | 0 |
| **All sites** | 676 | 1415 | 1345 | 923 | 2433 | 1040 |

Supplementary Table 5: Area Under the Curve (AUC) measured with cross-validation applied to the entire data set.

| **Splitting by Age/Sex** | | | | | | | | |
|---|---|---|---|---|---|---|---|---|
| | Cortical + Subcortical | | Cortical Thickness | | Cortical Surface area | | Subcortical Volume | |
| | No ComBat | With ComBat | No ComBat | With ComBat | No ComBat | With ComBat | No ComBat | With ComBat |
| **Elastic Net** | 0.649 | 0.524 | 0.616 | 0.499 | 0.595 | 0.518 | 0.634 | 0.524 |
| **LASSO** | 0.650 | 0.524 | 0.617 | 0.500 | 0.597 | 0.521 | 0.634 | 0.524 |
| **Ridge** | 0.648 | 0.524 | 0.615 | 0.500 | 0.596 | 0.517 | 0.634 | 0.524 |
| **SVM PCA** | 0.681 | 0.541 | 0.636 | 0.527 | 0.601 | 0.516 | 0.663 | 0.519 |
| **SVM + ttest** | 0.667 | 0.527 | 0.619 | 0.530 | 0.589 | 0.528 | 0.656 | 0.523 |
| **SVM linear** | 0.654 | 0.485 | 0.614 | 0.500 | 0.598 | 0.509 | 0.635 | 0.512 |
| **SVM rbf** | 0.677 | 0.536 | 0.636 | 0.530 | 0.610 | 0.513 | 0.664 | 0.528 |
| **Random Forests** | 0.676 | 0.549 | 0.633 | 0.515 | 0.615 | 0.522 | 0.662 | 0.532 |
| **Splitting by Site** | | | | | | | | |
| | Cortical + Subcortical | | Cortical Thickness | | Cortical Surface area | | Subcortical Volume | |
| | No ComBat | With ComBat | No ComBat | With ComBat | No ComBat | With ComBat | No ComBat | With ComBat |
| **Elastic Net** | 0.524 | 0.522 | 0.502 | 0.487 | 0.505 | 0.519 | 0.524 | 0.522 |
| **LASSO** | 0.524 | 0.523 | 0.492 | 0.487 | 0.507 | 0.520 | 0.524 | 0.522 |
| **Ridge** | 0.524 | 0.522 | 0.497 | 0.487 | 0.505 | 0.518 | 0.524 | 0.522 |
| **SVM PCA** | 0.524 | 0.526 | 0.494 | 0.510 | 0.498 | 0.528 | 0.525 | 0.526 |
| **SVM + ttest** | 0.510 | 0.521 | 0.484 | 0.504 | 0.509 | 0.514 | 0.510 | 0.528 |
| **SVM linear** | 0.523 | 0.509 | 0.499 | 0.507 | 0.506 | 0.507 | 0.517 | 0.525 |
| **SVM rbf** | 0.522 | 0.521 | 0.492 | 0.504 | 0.496 | 0.513 | 0.516 | 0.520 |
| **Random Forests** | 0.519 | 0.526 | 0.496 | 0.501 | 0.496 | 0.523 | 0.529 | 0.514 |

Supplementary Table 6: Sensitivity measured with cross-validation applied to the entire data set.

| **Splitting by Age/Sex** | | | | | | | | |
|---|---|---|---|---|---|---|---|---|
| | Cortical + Subcortical | | Cortical Thickness | | Cortical Surface area | | Subcortical Volume | |
| | No ComBat | With ComBat | No ComBat | With ComBat | No ComBat | With ComBat | No ComBat | With ComBat |
| **Elastic Net** | 0.581 | 0.514 | 0.544 | 0.488 | 0.533 | 0.508 | 0.572 | 0.510 |
| **LASSO** | 0.587 | 0.515 | 0.546 | 0.479 | 0.532 | 0.506 | 0.575 | 0.505 |
| **Ridge** | 0.582 | 0.513 | 0.548 | 0.489 | 0.535 | 0.510 | 0.573 | 0.509 |
| **SVM PCA** | 0.561 | 0.450 | 0.532 | 0.392 | 0.473 | 0.489 | 0.555 | 0.481 |
| **SVM + ttest** | 0.571 | 0.503 | 0.471 | 0.452 | 0.370 | 0.470 | 0.557 | 0.494 |
| **SVM linear** | 0.563 | 0.509 | 0.476 | 0.478 | 0.490 | 0.516 | 0.574 | 0.525 |
| **SVM rbf** | 0.576 | 0.413 | 0.516 | 0.452 | 0.499 | 0.487 | 0.555 | 0.476 |
| **Random Forests** | 0.415 | 0.080 | 0.442 | 0.237 | 0.336 | 0.126 | 0.429 | 0.145 |
| **Splitting by Site** | | | | | | | | |
| | Cortical + Subcortical | | Cortical Thickness | | Cortical Surface area | | Subcortical Volume | |
| | No ComBat | With ComBat | No ComBat | With ComBat | No ComBat | With ComBat | No ComBat | With ComBat |
| **Elastic Net** | 0.509 | 0.506 | 0.492 | 0.484 | 0.443 | 0.510 | 0.524 | 0.510 |
| **LASSO** | 0.507 | 0.508 | 0.498 | 0.484 | 0.436 | 0.513 | 0.525 | 0.507 |
| **Ridge** | 0.509 | 0.506 | 0.495 | 0.485 | 0.440 | 0.507 | 0.524 | 0.510 |
| **SVM PCA** | 0.441 | 0.491 | 0.458 | 0.453 | 0.360 | 0.529 | 0.445 | 0.516 |
| **SVM + ttest** | 0.463 | 0.517 | 0.381 | 0.426 | 0.278 | 0.484 | 0.456 | 0.502 |
| **SVM linear** | 0.477 | 0.506 | 0.433 | 0.479 | 0.395 | 0.508 | 0.492 | 0.522 |
| **SVM rbf** | 0.455 | 0.467 | 0.403 | 0.426 | 0.379 | 0.513 | 0.451 | 0.508 |
| **Random Forests** | 0.254 | 0.102 | 0.394 | 0.218 | 0.138 | 0.139 | 0.296 | 0.154 |

Supplementary Table 7: Specificity measured with cross-validation applied to the entire data set.

| Splitting by Age/Sex | | | | | | | | |
|---|---|---|---|---|---|---|---|---|
| | Cortical + Subcortical | | Cortical Thickness | | Cortical Surface area | | Subcortical Volume | |
| | No ComBat | With ComBat | No ComBat | With ComBat | No ComBat | With ComBat | No ComBat | With ComBat |
| **Elastic Net** | 0.637 | 0.532 | 0.623 | 0.513 | 0.604 | 0.526 | 0.614 | 0.531 |
| **LASSO** | 0.637 | 0.533 | 0.620 | 0.518 | 0.624 | 0.527 | 0.616 | 0.530 |
| **Ridge** | 0.637 | 0.533 | 0.622 | 0.507 | 0.612 | 0.520 | 0.614 | 0.531 |
| **SVM PCA** | 0.714 | 0.609 | 0.670 | 0.634 | 0.677 | 0.548 | 0.689 | 0.546 |
| **SVM + ttest** | 0.683 | 0.527 | 0.691 | 0.578 | 0.764 | 0.582 | 0.681 | 0.548 |
| **SVM linear** | 0.670 | 0.539 | 0.679 | 0.530 | 0.655 | 0.521 | 0.629 | 0.522 |
| **SVM rbf** | 0.703 | 0.636 | 0.684 | 0.578 | 0.656 | 0.533 | 0.682 | 0.550 |
| **Random Forests** | 0.811 | 0.934 | 0.744 | 0.791 | 0.811 | 0.893 | 0.794 | 0.877 |
| **Splitting by Site** | | | | | | | | |
| | Cortical + Subcortical | | Cortical Thickness | | Cortical Surface area | | Subcortical Volume | |
| | No ComBat | With ComBat | No ComBat | With ComBat | No ComBat | With ComBat | No ComBat | With ComBat |
| **Elastic Net** | 0.517 | 0.523 | 0.504 | 0.494 | 0.564 | 0.518 | 0.490 | 0.518 |
| **LASSO** | 0.518 | 0.525 | 0.484 | 0.494 | 0.580 | 0.514 | 0.488 | 0.517 |
| **Ridge** | 0.519 | 0.522 | 0.498 | 0.494 | 0.571 | 0.511 | 0.489 | 0.518 |
| **SVM PCA** | 0.612 | 0.550 | 0.546 | 0.571 | 0.649 | 0.520 | 0.595 | 0.525 |
| **SVM + ttest** | 0.541 | 0.508 | 0.593 | 0.571 | 0.736 | 0.533 | 0.564 | 0.552 |
| **SVM linear** | 0.546 | 0.533 | 0.563 | 0.510 | 0.603 | 0.505 | 0.519 | 0.520 |
| **SVM rbf** | 0.570 | 0.562 | 0.583 | 0.571 | 0.607 | 0.513 | 0.556 | 0.529 |
| **Random Forests** | 0.783 | 0.909 | 0.596 | 0.783 | 0.845 | 0.866 | 0.742 | 0.848 |

Supplementary Table 8: Balanced accuracy measured with cross-validation applied to the entire data set under different harmonization options.

| | Splitting by Age/Sex | | | Splitting by Site | | |
|---|---|---|---|---|---|---|
| | ComBat | ComBat-GAM | CovBat | ComBat | ComBat-GAM | CovBat |
| **Elastic Net** | 0.523 | 0.522 | 0.517 | 0.514 | 0.515 | 0.514 |
| **LASSO** | 0.524 | 0.523 | 0.517 | 0.517 | 0.513 | 0.514 |
| **Ridge** | 0.523 | 0.519 | 0.518 | 0.514 | 0.516 | 0.514 |
| **SVM PCA** | 0.529 | 0.521 | 0.523 | 0.520 | 0.520 | 0.528 |
| **SVM + ttest** | 0.515 | 0.503 | 0.513 | 0.512 | 0.505 | 0.512 |
| **SVM linear** | 0.524 | 0.526 | 0.521 | 0.519 | 0.520 | 0.518 |
| **SVM rbf** | 0.525 | 0.522 | 0.522 | 0.515 | 0.519 | 0.512 |

Supplementary Table 9: Balanced accuracy of support vector machines (SVM) with linear kernel without feature selection trained and validated with leave-one-site-out cross-validation. More extreme values of balanced accuracy are obtained for cohorts containing no healthy subjects. Note that ComBat brings these values closer to average across all cohorts.

| Name of site | No ComBat (Bacc) | With ComBat (Bacc) | Ratio MDD/HC |
|---|---|---|---|
| **SHIP T0** | 0.503017 | 0.50513 | 0.3373232 |
| **SHIP S2** | 0.474821 | 0.516075 | 0.3069977 |
| **StanfT1wAggr** | 0.535866 | 0.476998 | 0.9491525 |
| **Minnesota** | 0.532143 | 0.521429 | 1.75 |
| **CSAN** | 0.489966 | 0.531973 | 1.2244898 |
| **Jena** | 0.498268 | 0.549567 | 0.3896104 |
| **Calgary** | 0.522727 | 0.553147 | 1.0576923 |
| **Barc** | 0.489415 | 0.467742 | 1.9375 |
| **DCHS** | 0.514117 | 0.607013 | 0.295082 |
| **AFFDIS** | 0.496377 | 0.455534 | 0.7173913 |
| **Moraldilemma** | 0.67663 | 0.586051 | 0.5217391 |
| **FIDMAG** | 0.487815 | 0.535714 | 1.0294118 |
| **MPIP** | 0.532092 | 0.533309 | 1.5971564 |
| **ETPB** | 0.561086 | 0.529412 | 1.3076923 |
| **TIGER** | 0.569573 | 0.517625 | 4.4545455 |
| **AMC** | 0.313726 | 0.431373 | N/A |
| **Episca(Leiden)** | 0.563158 | 0.513158 | 0.6333333 |
| **SanRaffaele** | 0.911111 | 0.644444 | N/A |
| **MODECT** | 0.785714 | 0.309524 | N/A |
| **Gron** | 0.519048 | 0.585714 | 0.952381 |
| **CARDIFF** | 0.85 | 0.5 | N/A |
| **Singapore** | 0.434659 | 0.400568 | 1.375 |
| **StanfFAA** | 0.468254 | 0.373016 | 0.7777778 |
| **QTIM** | 0.506663 | 0.509183 | 0.3591549 |
| **Houst** | 0.491057 | 0.477512 | 0.5591398 |
| **Melb** | 0.568216 | 0.546551 | 1.4019608 |
| **NESDA** | 0.52952 | 0.533417 | 2.3692308 |
| **Socat_dep** | 0.527722 | 0.547468 | 0.79 |
| **UCSF** | 0.522197 | 0.506061 | 0.8522727 |
| **LOND** | 0.531005 | 0.609527 | 1.1311475 |
| **ALL SITES** | 0.513 | 0.515 | 0.74 |

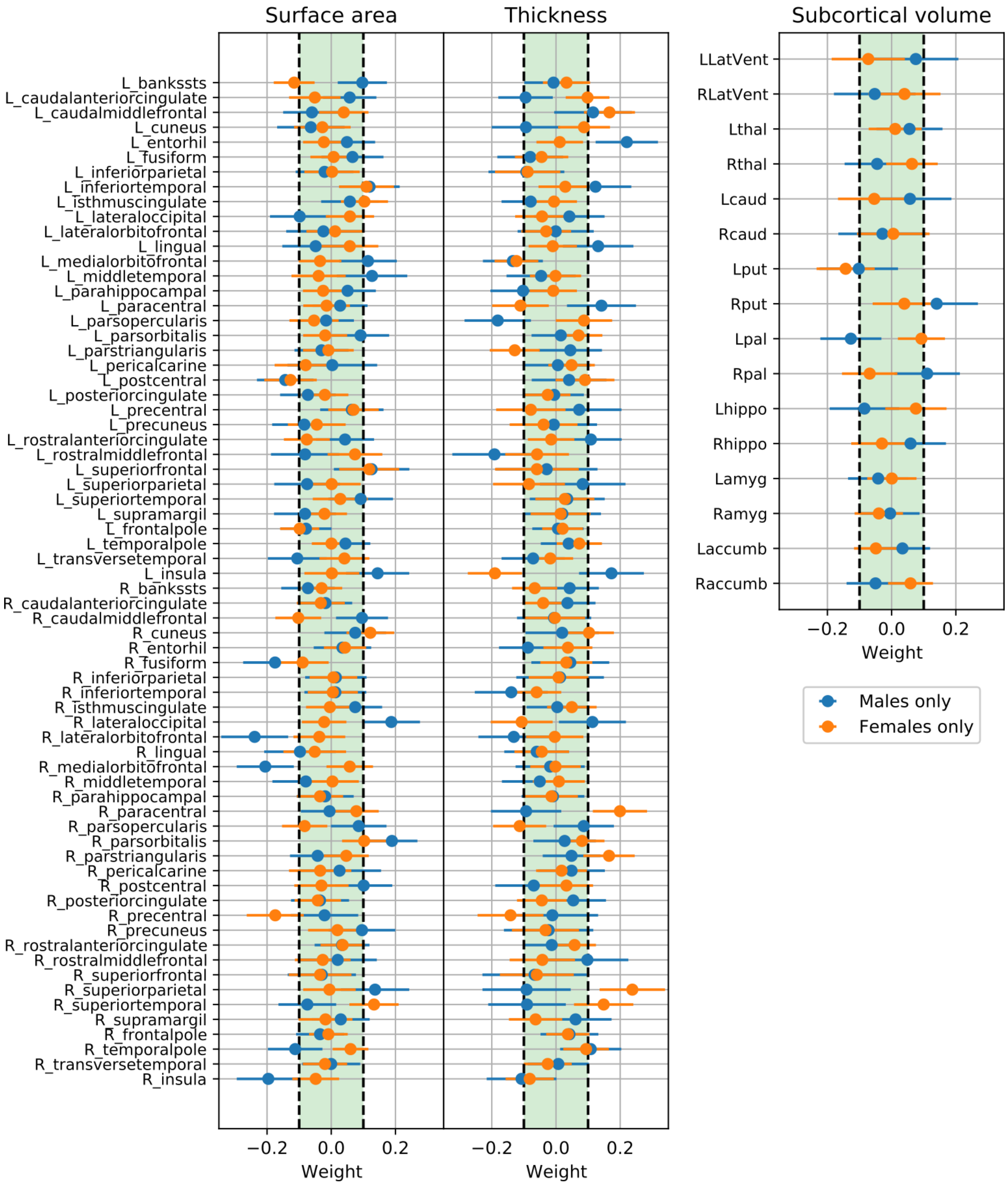

Supplementary Figure 1: Weights of SVM with linear kernel applied on stratified data by sex (no feature selection, with ComBat). The dashed lines represents ±0.1 as thresholds of feature weights, above which informative features stand out.

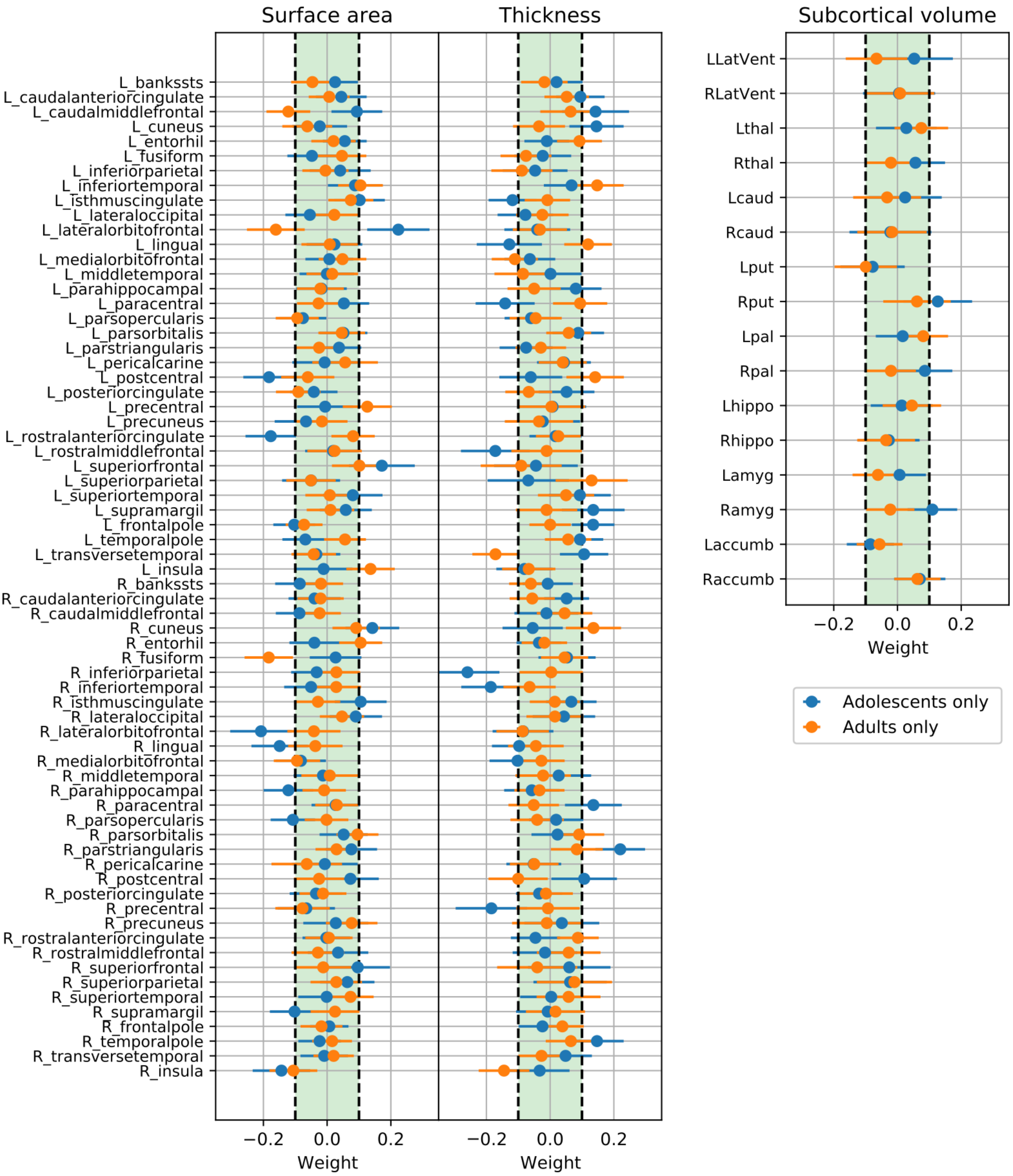

Supplementary Figure 2: Weights of SVM with linear kernel applied on stratified data by age of onset (no feature selection, with ComBat). The dashed lines represents ±0.1 as thresholds of feature weights, above which informative features stand out.

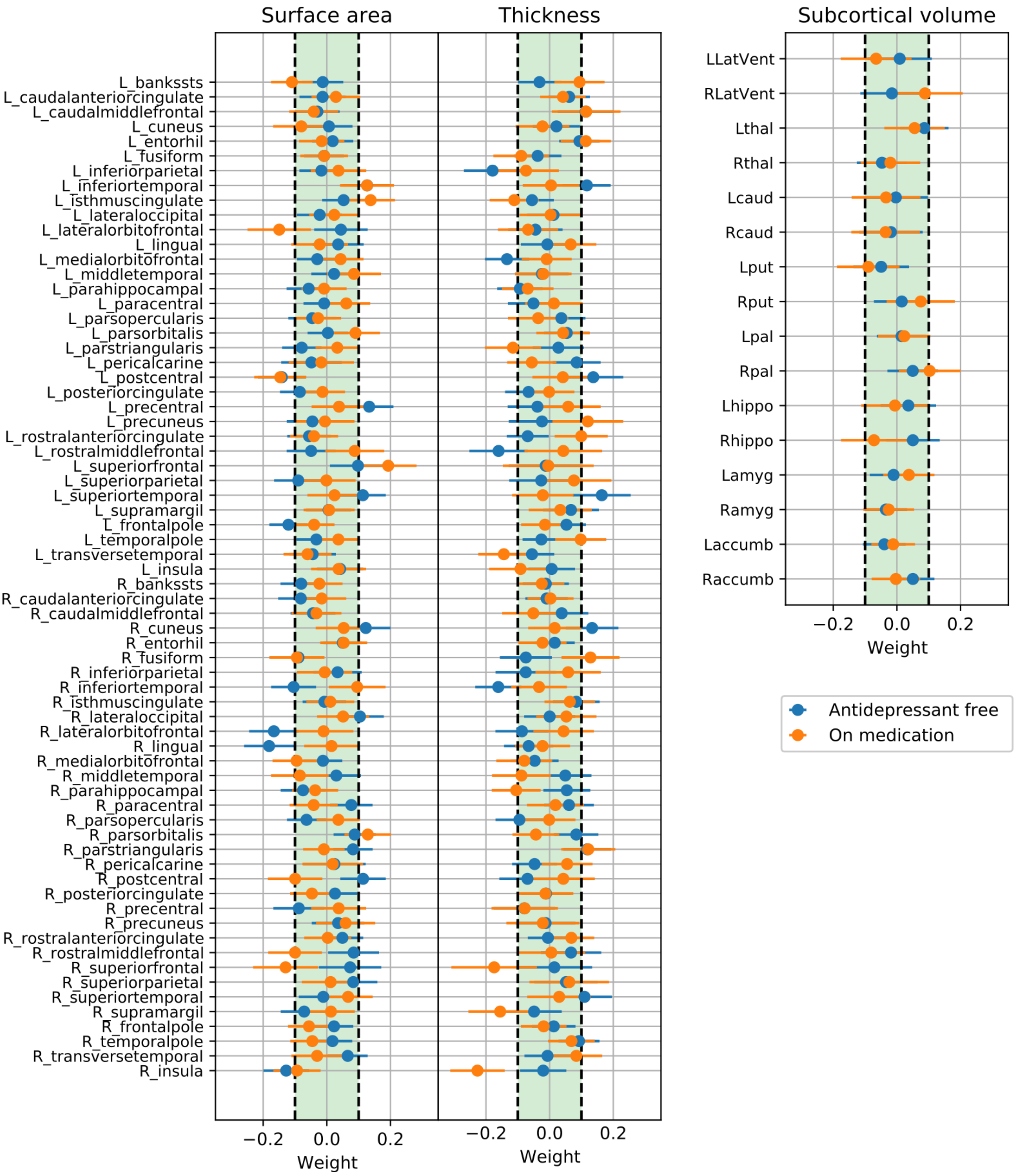

Supplementary Figure 3: Weights of SVM with linear kernel applied on stratified data by use of antidepressant medication (no feature selection, with ComBat). The dashed lines represents ±0.1 as thresholds of feature weights, above which informative features stand out.

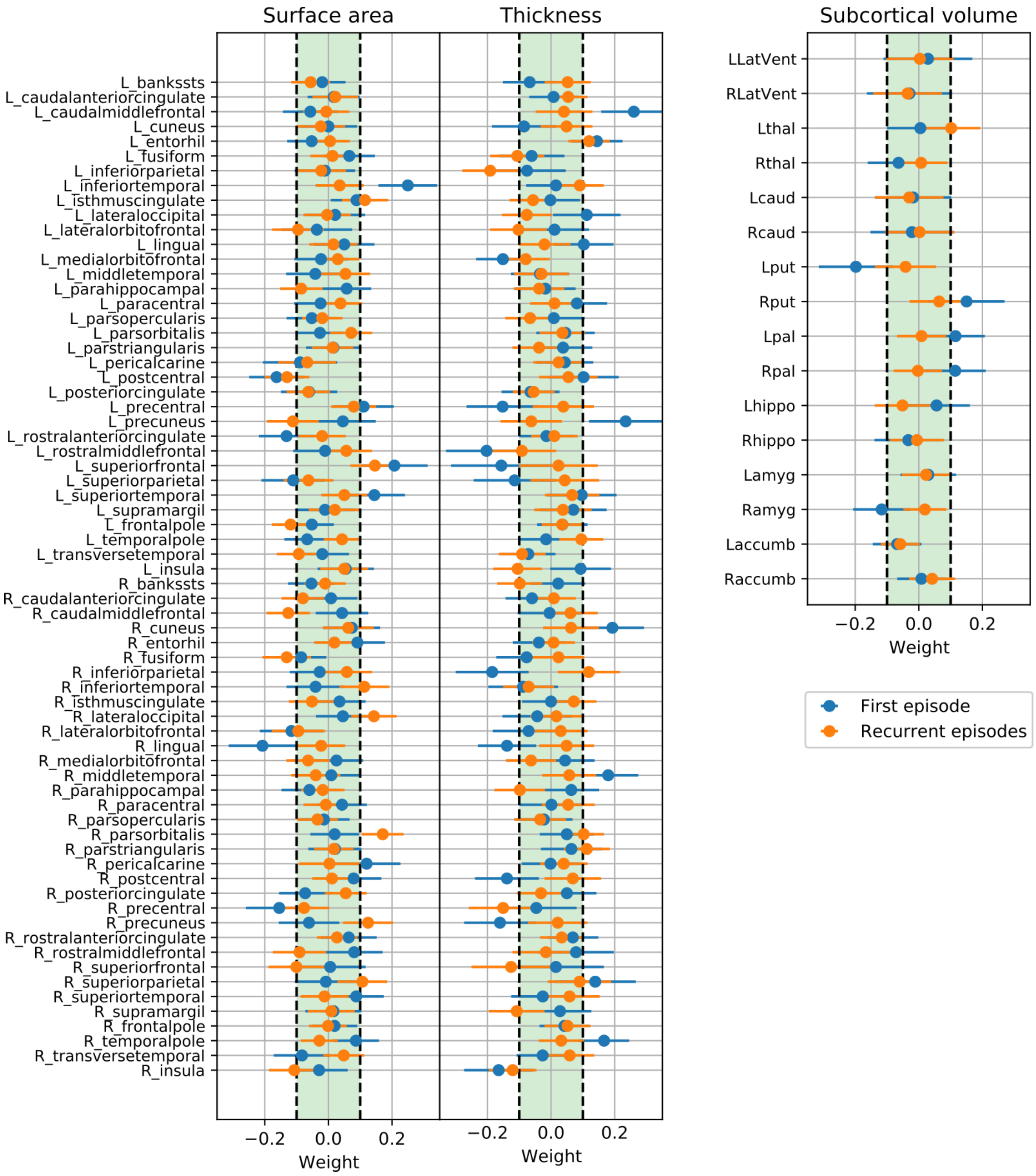

Supplementary Figure 4: Weights of SVM with linear kernel applied on stratified data by number of episodes (no feature selection, with ComBat). The dashed lines represents ±0.1 as thresholds of feature weights, above which informative features stand out.

*Harmonization methods*

We harmonized individual cortical and subcortical features by implementing the well-established statistical harmonization algorithm, ComBat [3]. Its purpose was to adjust Location (mean) and Scale (variation) (L/S) of all features of the data collected from different cohorts by preserving the influence of biologically-significant factors of interest in the features. Additionally, it is assumed that the site effect is independent across cortical and subcortical features. Additive and multiplicative site effects are assumed to be probabilistic and estimated by a Bayesian approach. In short, the empirical priors are added over the site specific means and variance [4]. Subsequently, the cortical and subcortical features would be standardized to mean of 0 and standard deviation of 1, while the site effect would be removed. ComBat assumes that the data $Y_{i,j,k}$ for ROI *k*, site *i* and subject *j* can be represented by the following model:

$$Y_{ijk} = \alpha_k + X_{ij}\beta_k + \gamma_{ik} + \delta_{ik}\varepsilon_{ijk} \quad (1)$$

Where $\alpha_k$ is an overall ROI value, $X$ is a design matrix where $X_{ij}$ is a vector containing site affiliation and controlled covariates of participant j in site i. In our case, these are age, sex and ICV. $\beta_k$ is the vector of regression coefficients corresponding to $X_{ij}$, $\gamma_{ik}$ and $\delta_{ik}$ correspond to additive and multiplicative site effect and $\varepsilon_{ijk}$ is an error term assumed to follow normal distribution with mean zero and variance $\sigma_k^2$. After parameter estimation in the model above, the standardized data $Y^*{}_{ijk}$ can be calculated as follows:

$$Y^*{}_{ijk} = \frac{Y_{ijk} - \hat{\alpha}_k - X_{ij}\hat{\beta}_k - \hat{\gamma}_{ik}}{\hat{\delta}_{ik}} + \hat{\alpha}_k + X_{ij}\hat{\beta}_k \quad (2)$$

where $\hat{\alpha}_k$, $\hat{\beta}_k$, $\hat{\gamma}_{ik}$ and $\hat{\delta}_k$ are estimated ComBat parameters. Additionally, it is assumed that the site effect is not independent across cortical and subcortical features.

All parameter estimations, which includes estimates of $\hat{\alpha}_k$, $\hat{\beta}_k$, $\hat{\gamma}_{ik}$ and $\hat{\delta}_k$, should be computed only on the training set, i.e. 9 CV folds, to avoid non-independence of the training and test data, also known as data leakage. After parameter estimations and training of the ML algorithm were complete, the calculated parameters were used to adjust the test data and the performance of the trained classification algorithm measured on the test set represented by the remaining CV fold. These parameters were directly used for adjusting data from unseen subjects from the test set only if these subjects belong to the same cohorts as in the training set. This scenario corresponds to Splitting by Age/Sex strategy as every CV fold contains subjects from all cohorts.

In Splitting by Site strategy, subjects from one cohort are included only in one CV fold, thus the direct usage of estimated ComBat parameters on the test set is imprudent. Here we adapted the approach of a reference batch adjustment [5], which constitutes fixing a reference sites, while other sites are adjusted to the mean and variance of the reference site according to the following equation:

$$Y^*_{ijkr} = \frac{Y_{ijk} - \hat{\alpha}_{rk} - X_{ij}\hat{\beta}_{kr} - \hat{\gamma}_{ikr}}{\hat{\delta}_{ikr}} + \hat{\alpha}_{kr} + X_{ij}\hat{\beta}_{kr} \quad (3)$$

where $\alpha_{kr}$, $\beta_{kr}$ correspond to coefficients estimated on the reference site *r*. Additionally, $\gamma_{ikr}$, $\delta_{ikr}$ represent additive and multiplicative differences between site i and r. In our case, the test set was adjusted to a unified batch made by integrating all cohorts from the training set and adjusting to common mean and variance by the ComBat, which allowed to harmonize unseen cohorts without data leakage from the training set to the test set (Supplementary Figure 5). Importantly, we assumed a common covariate model for MDD and HC during ComBat in both splitting strategies. A normative approach, i.e. estimating covariate regressors on HC only, is possible in case of Splitting by Age/Sex. However, in Splitting by Site, normative estimation of covariate factors in unseen sites would result in double dipping, as we would have to use information on diagnosis of the test subjects before estimating classification performance on them. Thus, to have a consistent comparison of splitting strategies, we did not apply normative covariate estimation in ComBat.

This framework was additionally extended to include non-linear preservation of the covariates by substituting $X_{ij}\beta_k$ with a Generalized Additive Model (ComBat-GAM) [6], allowing nonlinear age trends to be preserved during the harmonization step. Furthermore, we considered a CovBat model, which assumes an additional covariance site effect alongside with mean and variance corrections [7]. We tested ComBat's harmonization ability to remove the site effect from the full data by training cortical and subcortical features via SVM with a linear kernel to predict the site. This was mostly tested in cases where the number of sites was below 10 [6,8,9], so it is relevant in the context of our current analysis. We used the Splitting by Age/Sex strategy, since to predict site information from the test folds, this information should be presented in the training folds - and this would not be possible in the Splitting by Site strategy . The balanced accuracy was 0.854 before applying ComBat without correction for age, sex and ICV. Such a high performance confirmed the expected strong sites effect presented in the data, which may interfere with the main MDD vs HC classification task. The classification balanced accuracy dropped substantially to 0.031 after applying ComBat, indicating the significant removal of site-related information in cortical and subcortical features. Such a low accuracy comes from the fact that with ComBat site-related information was removed from the data, resulting in SVM always predicting SHIP_T0 – the biggest cohort. By assessing the confusion matrices we could evidence that - without the harmonization step - the classification algorithm was able to predict site affiliation of the subject successfully, except of sites coming from the same research group e.g., MPIP (two cohorts) and in case of SHIP_T0-SHIP_S2 and Melbourne-MoralDilemma pairs. As an illustration, an example of a feature being harmonized via ComBat may be seen in Supplementary Figure 6.

To further investigate whether differences across sites were mainly driven by irregular age and sex distributions, we repeated the classification task by also regressing out age and sex from the features. This resulted in 0.816 and 0.031 balanced accuracies for predicting site without and with ComBat harmonization respectively. By comparing the results with and without this residualization step, we could infer that the classification performance in differentiating sites only minor roots in age and sex distribution differences between sites. To see if the site effect was due to the differences in scanners and scan acquisition protocols between cohorts, we trained SVM to predict scanner type from cortical and subcortical features. The resulting accuracy was 0.875, even higher than only site prediction. This hints to the site effect being primarily caused by differences in acquisition equipment across sites.

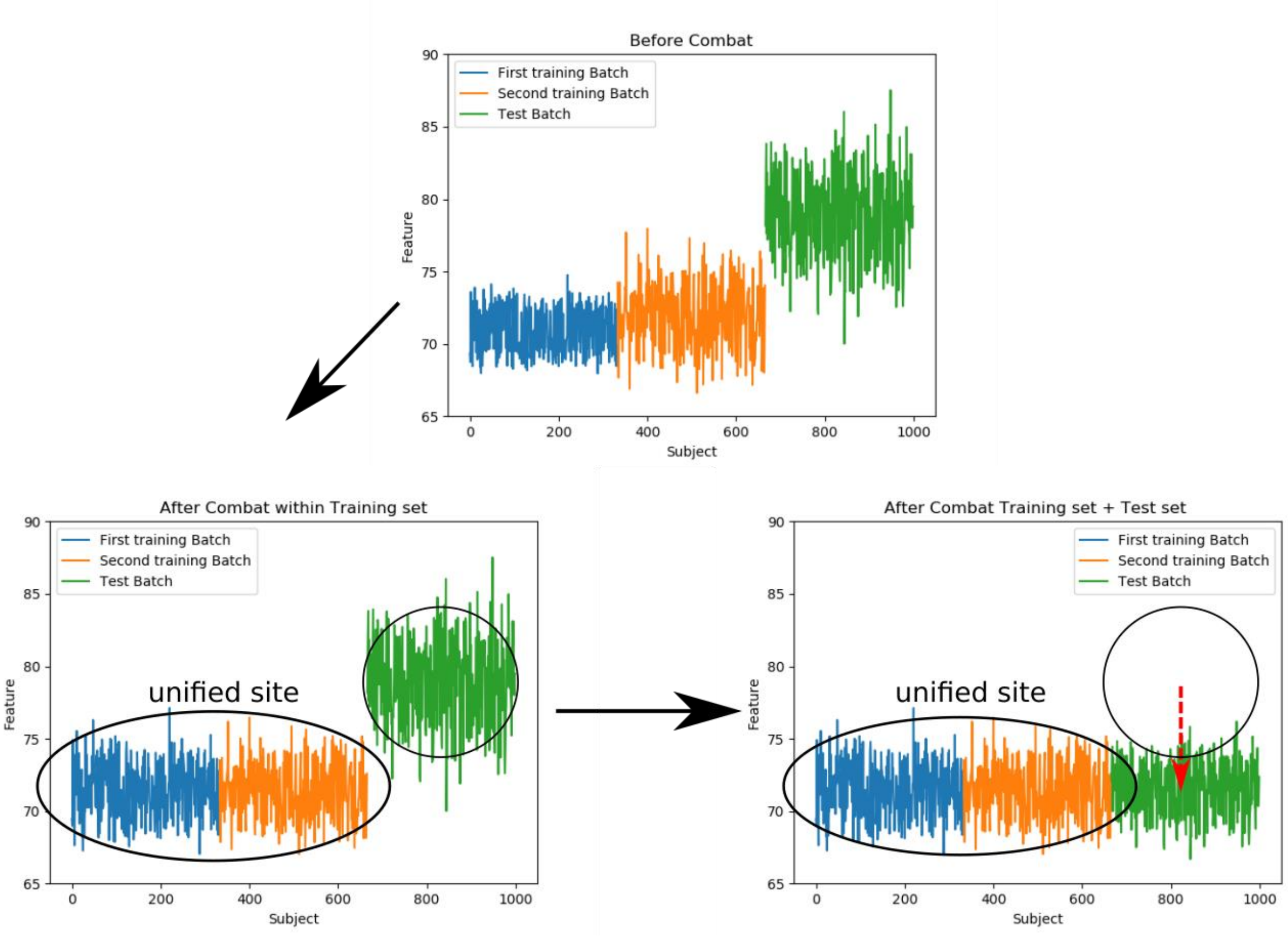

Supplementary Figure 5: Test set adjustment to the unified site. After ComBat is applied on the training set, all training sites are adjusted so that their residuals (after fitting covariates) have the same mean and variance, which we unify to build a unified site used for the classification training. After the training is complete, test set is harmonized to the fixed unified site allowing trained model to be evaluated on the test set.

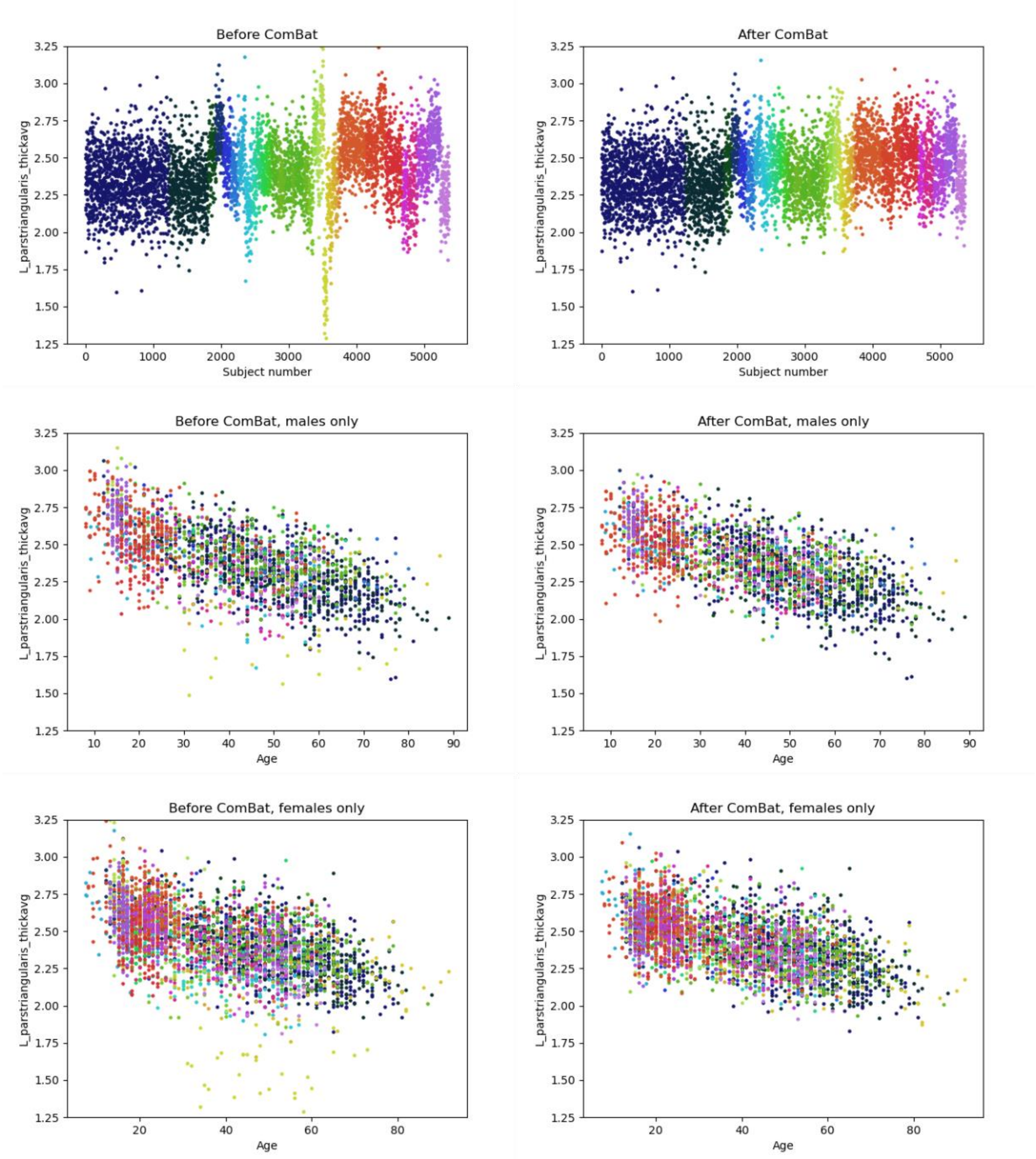

Supplementary Figure 6**:** An example of site effect removal by ComBat for left pars opercularis thickness. Color corresponds to the site affiliation. While the differences between sites are reduced, remaining differences correspond to age- and sex-related differences between cohorts (middle and bottom).